\documentclass[epj]{svjour}
\pdfoutput=1
\usepackage{graphicx}
\usepackage{upgreek}
\usepackage{multirow}
\usepackage{bm}        
\usepackage{amssymb}   
\usepackage{amsmath}   
\usepackage{xcolor}
\usepackage{microtype}
\begin{document}
\title{Towards a realistic setup for a dynamical measurement of deviations from Newton's $1/r^2$ law: the impact of air viscosity}

\author{J. Baeza-Ballesteros\inst{1,2},  A. Donini\inst{1,2}, G. Molina-Terriza\inst{3,4,5}, F. Monrabal\inst{3,4} and A. Sim\'on\inst{3,6}
}                     
\institute{Instituto de F\'{\i}sica Corpuscular, CSIC-Universitat de Val\`encia, 46980 Paterna, Spain \and Departamento de Física Teórica, Facultad de Física, Universitat de Val\`encia, 45100 Burjassot, Spain \and Donostia International Physics Center (DIPC), Paseo Manuel Lardizabal 4, Donostia-San Sebastian, E-20018, Spain \and Ikerbasque, Basque Foundation for Science, Bilbao, E-48013, Spain \and Centro de Física de Materiales, CSIC-UPV/EHU, Paseo Manuel de Lardizabal 5, 20018 Donostia, Spain \and Enrico Fermi Institute, University of Chicago, IL 60637, Chicago, USA}
\date{Received: date / Revised version: date}
%
\abstract{
A novel experimental setup to measure deviations from the $1/r^2$ distance dependence of Newtonian gravity was proposed in Ref.~\cite{Donini:2016kgu}. The underlying theoretical idea was to study the orbits of a microscopically-sized planetary system composed of a ``Satellite'', with mass $m_{\rm S} \sim {\cal O}(10^{-9})$ g, and  a ``Planet'', with mass $M_{\rm  P}  \sim {\cal O} (10^{-5}) $ g at an initial distance of hundreds of microns. 
The detection of precession of the orbit in this system would be an unambiguous indication of a central potential with terms that scale with the distance differently from $1/r$. This is a huge advantage with respect to the
measurement of the absolute strength of the attraction between two bodies, as most electrically-induced background potentials do indeed scale as $1/r$. Detection of orbit precession is unaffected by these effects, allowing for better sensitivities.  In Ref.~\cite{Baeza-Ballesteros:2021tha}, 
the impact of other subleading backgrounds that may induce orbit precession, such as, {\em e.g.}, the electrical Casimir force or general relativity, was studied in detail. It was found that the proposed setup could test Yukawa-like corrections, $\alpha \times \exp(-r/\lambda)$, to the $1/r$ potential with couplings as low as $\alpha \sim 10^{-2}$ for distances as small as $\lambda \sim 10$  $\upmu$\textup{m}, improving by roughly an order of magnitude present bounds. In this paper, we start to move from a theoretical study of the proposal to a more realistic implementation of the experimental setup. As a first step, we study the impact of air viscosity on the proposed setup and see how the setup should be modified in order to preserve the theoretical sensitivity achieved in Refs.~\cite{Donini:2016kgu,Baeza-Ballesteros:2021tha}. 
   } 

\authorrunning{J. Baeza-Ballesteros et al.}
\titlerunning{A realistic setup for a dynamical measurement of deviations from Newton's law}

\maketitle

\section{Introduction}
\label{sec:intro}

The discovery of the Higgs boson at the LHC  \cite{Aad:2012tfa} is the last step in the construction of a consistent interpretation of strong, weak and electromagnetic interactions, the Standard Model (SM). Much of the present effort at the LHC and other facilities is devoted, to either confirm or put to test predictions of the SM, something that up to the moment has been extremely successful. 
A strong consensus exists, though, that the SM is not the ``ultimate'' theory, but just a low-energy effective theory 
that must be extended to a more fundamental one at energies higher than those currently tested. Several experimental observations cannot be explained within the boundaries of the SM: a huge amount of data suggests the existence of matter that gravitates but does not emit light 
(called, unambiguously, Dark Matter); something (called Dark Energy) is held responsible for the accelerated expansion of the Universe;
even if the SM conserve baryon number, there is no significant amount of anti-matter in the Universe; and, we have no unique model
to explain neutrino masses  (the most compelling evidence for physics beyond the SM so far).
In addition to these experimental evidences, two long-standing theoretical problems arise above all: the so-called ``hierarchy problem" and the
quantization of gravity. 

If the SM is indeed an effective theory, then masses of SM particles are expected to be ``naturally" close to the scale at which the SM should be replaced by a more fundamental theory \cite{'tHooft:1979bh}, although lighter particles should be protected by the recovery of some symmetry. This is indeed the case for the SM gauge bosons, that are either massless or with a mass 
proportional to the electro-weak symmetry breaking scale $\Lambda_{\rm EW}$, the scale at which the gauge group $\text{SU}_\text{c} (3) \times \text{U}_Q(1)$
is replaced by $\text{SU}_\text{c}(3) \times \text{SU}_\text{L} (2) \times \text{U}_Y(1)$. Even though one may claim that the restoring of chiral symmetry, or of some yet unknown flavor symmetry, may account for the large spread in fermion masses, the very existence of the Higgs boson poses a new theoretical problem: its mass, $ m_{\rm H}=125.10\pm 0.14$ GeV \cite{Zyla:2020zbs}, could be considered ``natural'' according to the 't Hooft naturalness criterion. However, loop corrections tell us that scalar particles have the very peculiar feature that their masses are quadratically sensitive to the cut-off scale $\Lambda_{\rm UV}$ at which the effective theory is no longer valid. This means that $m_H$ should be sensitive to a scale much higher than $\mathit{\Lambda}_{\rm EW}$, contrary to the observed experimental value. This is indeed the aforementioned ``hierarchy problem". 

Several proposals for a model ``more fundamental" than the SM have been advanced over the years, the main hypothesis being that New Physics is ``just around the corner" of the TeV, thus implicitly enhancing the discovery potential of the LHC. The observed Higgs mass points precisely at such a theory, one in which quadratically divergent quantum corrections going as $ c_{\rm W} \Lambda^2_{\rm UV}$ are small due to the Wilson coefficient $c_{\rm W}$ being ${\cal O} \left (1 / \Lambda^2_{\rm UV} \right )$. If $\Lambda_{\rm UV}$ is larger (but not much larger) than $\Lambda_{\rm EW}$, then
a moderate fine-tuning of the Wilson coefficient could protect the observed value of $m_{\rm H}$ against quantum corrections.
This is the case of the Minimal Supersymmetric Standard Model \cite{Dimopoulos:1981zb} and of Technicolor in its different disguises
\cite{Susskind:1978ms,Farhi:1980xs,Belyaev:2013ida}, for example. Non-observation of supersymmetric particles or of known particles compositeness (or of any other kind of new physics at the LHC) has indeed unveiled our huge miscomprehension of the mechanisms that lie behind the very foundation of the Standard Model. Is it indeed an effective theory? And, if not, how can the experimental evidences
of physics beyond the SM be addressed? 

Among the proposals to explain the smallness of the observed Higgs mass with respect to a more fundamental scale, one stands on its own: the idea that gravity may propagate across more than three spatial dimensions, being the (so-called) {\em extra-dimensions} either compact and tiny or with significant curvature. By compactifiying the extra-dimension, it can be shown that
the fundamental scale of gravity may differ significantly from the Planck mass, $M_{\rm Pl}\sim10^{19}$~GeV, 
or that extra-dimensional gravity effectively couples with the Higgs at a much smaller scale, thus explaining the hierarchy problem. Common to all of these models is that gravity is modified at distances smaller than some length scale $\lambda_{\rm ED}$. Notice that
the scale below which this occurs may differ enormously between any two different extra-dimensional models: for large extra-dimensions (LED) models \cite{Antoniadis:1990ew,Antoniadis:1998ig} in order to have the actual fundamental scale of gravity slightly larger than the TeV, we need $\lambda_{\rm ED} \in [10^{-3},1]$ mm, depending on the number of extra-dimensions; for the Randall-Sundrum (RS) model \cite{Randall:1999ee,Randall:1999vf}, $\lambda_{\rm ED}$ is typically of the order of $10^{-27}$ mm; and for the clockwork/linear dilation (CW/LD) model 
\cite{Giudice:2016yja,Giudice:2017fmj}, $\lambda_{\rm ED}$ lies in between these two extremes. The need for a new theory of gravity below $\lambda_{\rm ED}$ is, {\em per se}, an interesting requirement of these models, as it may represent a first step on the path of a consistent quantum theory of gravity, a long-standing theoretical problem for which we have no clue since the first half of the last century. 

When weak gravitational fields are considered, constraints on anomalous behavior of gravity at short distances may be obtained
by looking for deviations from the Newtonian force between two test masses (see, {\em e.g.}, Ref.~\cite{Buisseret:2007qd}). 
A summary of relatively recent experimental bounds can be found in Ref.~\cite{Adelberger:2009zz} (where the results of different techniques from Refs.~\cite{Hoyle:2004,Kapner:2006si,Spero:1980,Hoskins:1985,Tu:2007,Long:2003,Chiaverini:2003,Smullin:2005} are shown together), 
with the most recent results published in Ref.~\cite{Lee:2020zjt}. 
Other results can also be found in  Refs.~\cite{Perivolaropoulos:2016ucs,Antoniou:2017mhs,Perivolaropoulos:2019vkb}.
Although these constraints have been obtained using very different techniques, they all rely on the measurement of 
the absolute strength of the gravitational force acting between two bodies at a given distance $r$. The shorter the distance, 
the larger the gravitational attraction, but also, the larger the unavoidable electrically-induced forces 
between the two bodies.  Bodies with non-negligible electric charge feel Coulombian attraction or repulsion. However, even in the case in which the two bodies are electrically ``neutral" ({\em i.e.} with a very small residual charge), inhomogeneous charge distribution within the bodies may
induce Van der Waals and dipolar forces that will act between them. A common feature to all these electrically-induced background potentials is that
they can have a $\sim 1/r$ dependence with the distance between the two bodies, {\em i.e.} the same dependence of the Newtonian
gravitational potential. This means that the main obstacle to improve the sensitivity of these experiments is the need to distinguish
the measurement of an ``effective" gravitational constant $G_{\rm N}^\prime = G_{\rm N} \, (1 + B)$ (that includes the backgrounds $B$) from the true interaction $G_{\rm N}$, 
to test if the Newtonian gravitational attraction at short distances is indeed the same as at large distances.

From a phenomenological point of view, it is useful to parametrize deviations from the $1/r$ Newtonian dependence in terms of a Yukawa potential, $\delta V \propto \alpha/r \times \exp(-r/\lambda)$, where $\alpha$ is an effective coupling and $\lambda$ the characteristic
scale at which New Physics starts to be relevant. For $\alpha = 1$, the bound obtained in Ref.~\cite{Lee:2020zjt} is $\lambda < 38.6$ $\upmu$\textup{m} at 95$\%$ confidence level (CL). This bound may be immediately applied to LED models, as in this particular extension of the SM the typical values
of $\alpha$ are proportional to twice the number of extra-dimensions 
({\em i.e.} $\alpha \in [2,12]$ for $n \in [1,6]$). The bound is, of course, irrelevant for the RS model, for which $\lambda \propto M_{\rm P}^{-1}$, whereas for the CW/LD model this bound may constrain some portion of the parameter space. Other values of $\alpha$ are characteristic of other extensions of the SM, such as the so-called fifth-force \cite{Berge:2021yye}, or quintessence \cite{Zlatev:1998tr,Yang:2018xah}. In order to improve our present bounds, it is therefore of great interest  to look for new methods 
that may allow to bypass the problems related to the electrical backgrounds and, thus, break the barrier of the tens of microns. 

A proposal to attain this goal was advanced in Ref.~\cite{Donini:2016kgu}, where it was suggested that the geometry of the trajectory of a microscopical test body {\em orbiting} around a heavier one that acts as the source of gravitational field may avoid most of the electrically-induced backgrounds.
The motivation is that the electrically-induced corrections to the Newtonian potential scale with the distance as $1/r$, whereas a Yukawa potential
has a different $r$-dependence. According to Bertrand's theorem (see, {\em e.g.}, Ref.~\cite{romero1997contemporary}), this term induces precession
of the orbit, whereas $1/r$ effects do not. Therefore, {\bf precession of the orbit} would be a smoking gun for deviations from the $1/r$-dependence of the Newtonian potential independent of (possibly dominant) electrically-induced backgrounds. We  studied the sensitivity of a planetary system consisting
of a Planet of mass $M_{\rm P} \sim 10^{-5}$ g and a Satellite of mass $m_{\rm S} \in [10^{-9},10^{-8}]$ g. The proposed Planet would be made of platinum (so as to be the smallest as possible at room temperature) and the Satellite of pyrolytic graphite, a diamagnetic material that would allow to levitate it with an ${\cal O}(1)$ T commercial magnets in order to cancel the effect of Earth's gravitational attraction. 
In the proposed experimental setup, the Satellite is put into motion by means of photo-irradiation 
with initial conditions chosen as to put it into a bounded orbit around the Planet, with an initial distance between the centers of the two bodies of $r_0 \in [100,200]$ $\upmu$\textup{m}.
The theoretical sensitivity to deviations from the $1/r^2$ Newton's law of such a setup was found to reach $\lambda < 10$ $\upmu$\textup{m} for $\alpha = 1$ at 95\% CL. Subleading backgrounds, that do induce precession of the orbit, were later studied in Ref.~\cite{Baeza-Ballesteros:2021tha}, where
it was shown that the proposed experimental setup should be able to improve present bounds on $\lambda$ by a factor 5 to 10 for any value of $\alpha \in [5\times10^{-3}, 10^3]$, the range of the 
 improvement depending significantly on sub-leading background terms scaling as $1/r^2$. 
 
 It is now time to take a step forward and try to move from an ``gedanken" (albeit realistic) setup to a ``realistic" one. This is what we plan to start doing in this paper, by taking into account the
 impact that air viscosity may have on our setup and modify it accordingly in order to maintain the goal sensitivity to deviations from the Newton's law. 
 The paper is organized as follows: in Sect.~\ref{sec:theo} we remind the very simple classical mechanics that we are going to use throughout the paper and the state-of-art of the would-be sensitivity as of Refs.~\cite{Donini:2016kgu,Baeza-Ballesteros:2021tha}; 
in Sect.~\ref{sec:viscosity} we study the impact of air viscosity on the orbit of the Satellite when moving around the Planet, and compute the vacuum required in order to preserve the sensitivity of the setup; 
in Sect.~\ref{sec:setup} we eventually present a ``realistic" experimental setup to deal with possible backgrounds and, specially, with air viscosity; 
in Sect.~\ref{sec:sens} we compute the expected sensitivity of this updated setup to New Physics;  and in Sect.~\ref{sec:concl}, we eventually come to a conclusion.

\section{Theoretical framework and previous work}
\label{sec:theo}

Consider a gravitational system consisting of two spheres, called the ``Planet" (P) and the ``Satellite" (S). 
The mass of former, $M_{\rm P}$, is taken to be much larger than that of the latter,  $m_{\rm S}$, so that the motion of the Planet under the effect of the Satellite can be safely neglected.\footnote{The motion of the Planet due to the gravitational attraction of the Satellite can be easily included when numerically solving the equations of motion of the system, but its effect has been checked to be negligible for the masses considered.} Thus, we consider the Planet at rest at the origin of coordinates and let $\bm{r}(t)$ be the position of the Satellite 
at time $t$.

The equation of motion for the Satellite is:
\begin{equation}
\label{eq:newtonforce}
m_{\rm S} \ddot{{\bm{r}}} = {\bm{ F}} (\bm{r}) = - \bm{ \nabla} V (r)\,, 
\end{equation}
where $V(r)$ is the gravitational potential with $r=|\bm{r}|$. In the Newtonian case, it takes the form $V_\text{N}(r)=-G_\text{N}/r$, with $G_\text{N}$ the Newtonian constant of gravitation. The orbit of the Satellite is a conic section characterized by its eccentricity, $e$, and initial angular momentum, $h_0$. Here, we focus on the case of a bounded system, and so consider $e<1$, for which the orbit is a closed ellipsis (or circle if $e=0$).

Deviations from the Newtonian case, called ``Beyond-Newtonian'' (BN), are usually represented in terms of a Yukawa potential, 
\begin{equation}
\label{eq:YukawaPotential}
V_{\rm BN} (r) = - \frac{G_{\rm N} M_{\rm P}}{r} \, \left [ 1 + \alpha \, \text{e}^{-r/\lambda} \right]  \, ,
\end{equation}
with $\alpha$ a dimensionless coupling and $\lambda$ the length scale below which New Physics becomes relevant, so that for $r \gg \lambda$, such effects are exponentially suppressed. The orbit of the Satellite in the presence of such New Physics is very different to the Newtonian case. According to Bertrand's theorem, closed orbits are only possible for central potentials with a radial dependence of the form $1/r$ or $r^2$, and deviations from these two options imply that the orbit is not closed, but instead precedes around the Planet. For the BN potential, the direction of the precession depends on the sign of $\alpha$. In this work, as it was done in Refs.~\cite{Donini:1999px,Baeza-Ballesteros:2021tha}, only $\alpha \geq 0$ is considered, for which we expect precession in the same direction as the movement of the Satellite, with an extension to $\alpha<0$ being straightforward. 

In Ref. \cite{Donini:2016kgu},  an experimental setup to test possible deviations to Newtonian gravity exploiting this effect was proposed, consisting of a Planet of mass $M_\text{P}\sim10^{-5}$ g and a Satellite with mass $m_\text{S}\in[10^{-9},10^{-8}]$ g. This experiment, whose goal is to test the dynamical properties of the system, has the advantage over static experiments that its sensitivity is not affected by leading electromagnetic backgrounds, as the electrostatic potential goes as $1/r$ and so does not induce precession. By measuring the orbit along several periods (defined as the time it takes to the Satellite to perform a $2\pi$-orbit around the Planet) and comparing the results with those expected in case
of a Newtonian orbit, it was shown that we could improve current bounds on $\lambda$ by up to an order of magnitude for $\alpha\in[5\times10^{-3},10^3]$.

As a subsequent step, Ref.~\cite{Baeza-Ballesteros:2021tha} optimized the experimental setup and analyzed the effect of possible backgrounds. These were parametrized as an expansion in $1/r$ using the following modified-Yukawa (mY) potential
\begin{eqnarray}\
\label{eq:BNPotential}
V_{\rm mY}(r) &=& -\frac{G_{\rm N} \, M_{\rm P}}{r} \, \left( 1 + Q_2 + \frac{Q_3}{r} + \frac{Q_4}{r^2}  + \dots\right . \nonumber \\
&+& \left . \alpha \, e^{-r/\lambda} \right ) \, , 
\end{eqnarray}
where $Q_k$ are the coefficients of a force whose $r$-dependence is proportional to $1/r^k$. Note that only the $Q_k \geq 0$ case was considered: in the case of positive $\alpha$, such backgrounds would be the most impactful on the experimental sensitivity, as they will induce precession in the same direction as the Yukawa term. The mY potential is expected to account for the effect of different background sources, including multipolar and van der Waals electrostatic forces, the electric Casimir effect or general relativity, among others. We note that the leading background, $Q_2$, which is expected to be related to electric effects, would not induce precession and so does not spoil our sensitivity. For the other $Q_k$, we found $Q_3$ to have the largest impact, with $Q_4$ having a much more reduced effect. We also found that the effect of 
$Q_{k\geq 4}$ is negligible.

In Ref.~\cite{Baeza-Ballesteros:2021tha} we optimized the setup and considered a Planet with $M_\text{P}=0.75\times10^{-5}$ g made of platinum and a Satellite made of pyrolytic graphite with $m_\text{S}=1.2\times10^{-9}$ g, which could be levitated using commercial neodymium magnets. Two different initial conditions were considered, characterized by different initial angular and radial velocities, $\dot{\theta}$ and $\dot{r}$, respectively: 
\begin{itemize}
\item {\bf Case 1}: $r_0=111.8$ $\upmu$\textup{m}, $\dot{r}_0=30.6$ nm s${}^{-1}$ and $\dot{\theta}_0=491.1$ $\upmu$rad s${}^{-1}$; 
          for which we have an apoapsis distance $r_{\rm a}\sim 150$ $\upmu$\textup{m} and a Newtonian period $T_{\rm N}\sim2$  h 30 min.
\item {\bf Case 2}: $r_0=177.7$ $\upmu$\textup{m}, $\dot{r}_0=13.4$ nm s${}^{-1}$ and $\dot{\theta}_0=259.2$ $\upmu$rad s${}^{-1}$; 
         for which we have $r_{\rm a}\sim 200$ $\upmu$\textup{m}, $T_{\rm N}\sim4$ h 30 min.
\end{itemize}
For both cases, the trajectory of the Satellite was simulated for 30 revolutions, and the maximum shift of the period was used as an observable. 

\begin{figure*}[h]
    \centering
    \begin{minipage}{0.45\textwidth}
        \centering
        \includegraphics[width=\textwidth, trim={2.5cm  0cm 1.5cm 0cm}]{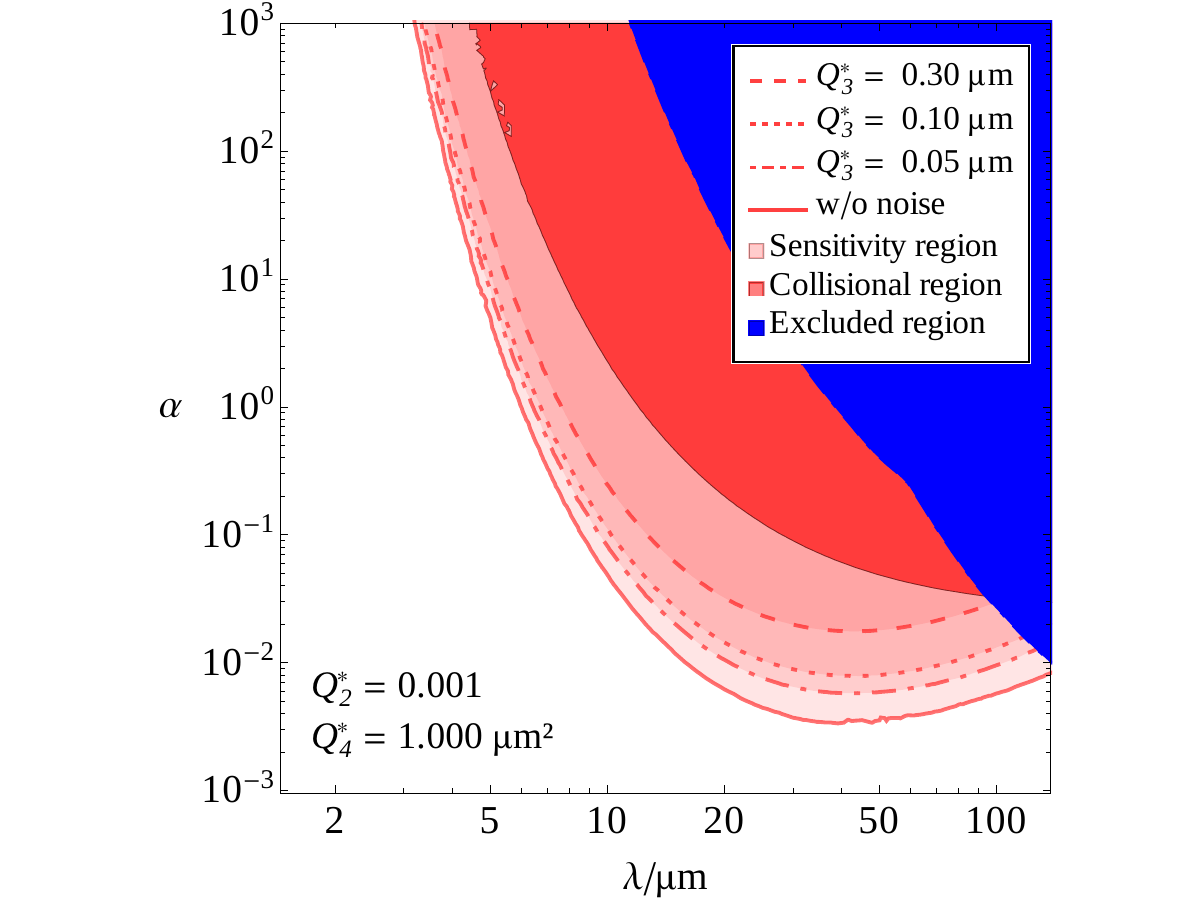} 
    \end{minipage}\hspace{1cm}
    \begin{minipage}{0.45\textwidth}
        \centering
        \includegraphics[width=\textwidth, trim={2.5cm 0cm 1.5cm  0cm}]{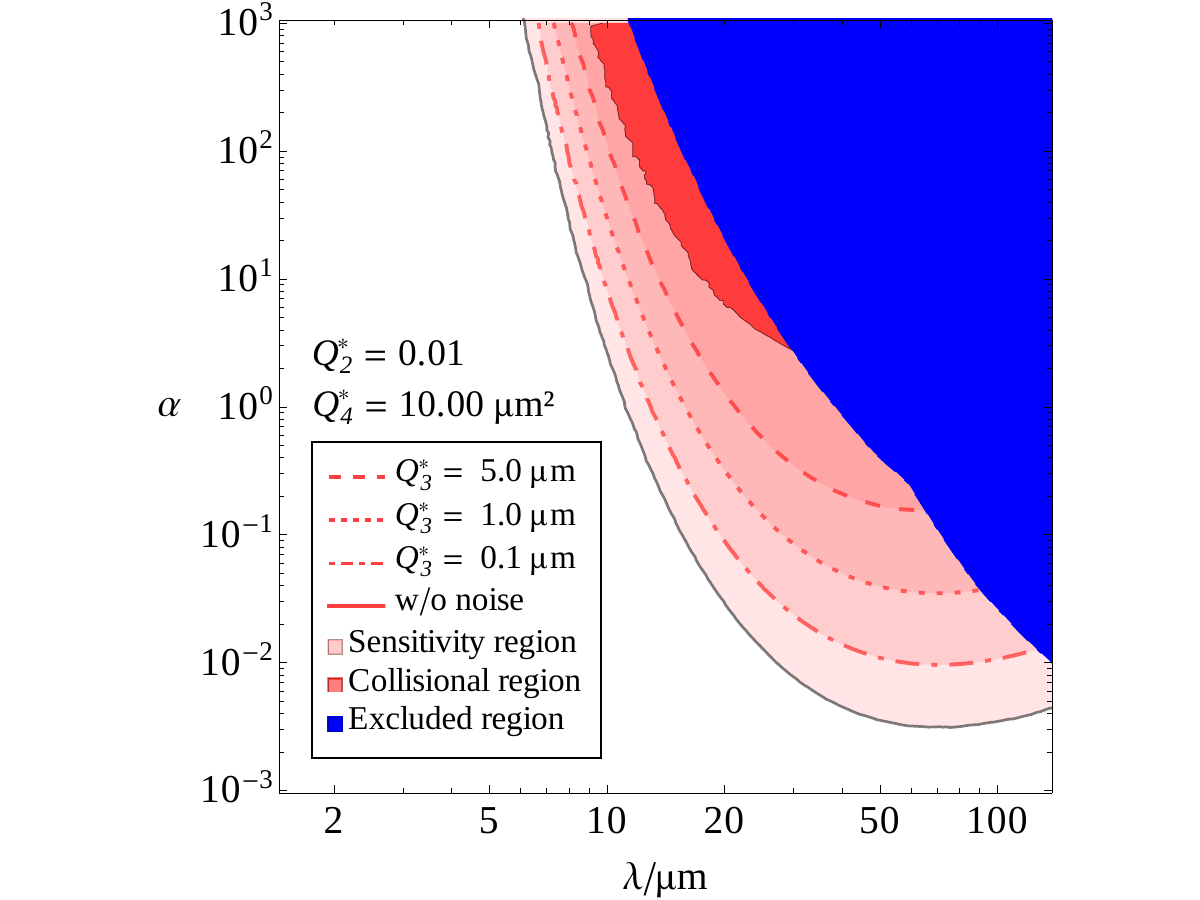} 
    \end{minipage}

    \caption{
    \it
    The 95 \% CL sensitivity of the proposed experimental setup in the ($\lambda,\alpha$) plane in the presence of attractive $Q_{k\leq 4}$ backgrounds smaller than some upper limit, $Q_k^*$, from Ref.~\cite{Baeza-Ballesteros:2021tha}. 
    The dark blue region represents the present experimental bounds,
    from Ref.~\cite{Lee:2020zjt}. 
    The red region is the part of the parameter space for which a Beyond-Newtonian potential will induce collision between \textup{S} and \textup{P}. The light-shaded pink region is the part of the parameter space in which we should detect precession due to New Physics large enough to be distinguishable from that induced by the backgrounds. Left panel: {\bf Case 1}.  $Q_2^* = 10^{-3}$, $Q_4^*= 1$ $\upmu\textup{m}^2$; $Q_3^* = 0.05$ $\upmu$\textup{m} (dot-dashed line), $Q_3^* = 0.1$ $\upmu$\textup{m} (dotted line), and $Q_3^* = 0.5$ $\upmu$\textup{m} (dashed line), as reported in the plot legend.
    Right panel: {\bf Case 2}. $Q_2^* = 10^{-2}$, $Q_4^*= 10$ $\upmu\textup{m}^2$; $Q_3^* = 0.1$ $\upmu$\textup{m} (dot-dashed line), $Q_3^* = 1.0$ $\upmu$\textup{m} (dotted line), and $Q_3^* = 5.0$ $\upmu$\textup{m} (dashed line). 
    In both panels the solid line corresponds to the noiseless scenario, $Q_k = 0$.
    }
        \label{fig:Limits2}
\end{figure*}

In Fig.~\ref{fig:Limits2}, we reproduce some of the main results of Ref.~\cite{Baeza-Ballesteros:2021tha}, which show  the experimental sensitivity at the 95 \% CL. This sensitivity is defined by comparing the signal generated with eq.~(\ref{eq:YukawaPotential}) with that coming from eq.~(\ref{eq:BNPotential}) in the case $\alpha=0$ (no New Physics). Each panel corresponds to one of the two aforementioned Cases, with the blue region being the excluded region from Ref.~\cite{Lee:2020zjt}, the dark red region the \textit{collisional region} in which the Satellite collides into the Planet in the presence of New Physics, and the light pink regions correspond to different choices of the nuisance parameters. {\bf Case 1} (left panel) gives a larger sensitivity, but it allows only for very small backgrounds, while in {\bf Case 2} (right panel) we can have bigger backgrounds at the expense of some sensitivity loss.

The computations performed in Refs.~\cite{Donini:2016kgu,Baeza-Ballesteros:2021tha}, however, correspond to a ``gedanken'' experiment and were carried out under different assumptions. It was considered that the experiment took place on an  empty volume and so the Satellite moved around the Planet in vacuum, but the presence of some air is unavoidable.  Also, in Ref.~\cite{Baeza-Ballesteros:2021tha}, we followed the orbit for 30 revolutions, which corresponds to more than two days of data gathering during which the setup needs to be kept stable, which is unrealistic. In this work, we move towards a more realistic setup, by considering the effects of the air viscosity, that cannot be parametrized with the backgrounds of eq.~\ref{eq:BNPotential}, and also by proposing a new observable based on the position of the Satellite.

\section{Impact of air viscosity}
\label{sec:viscosity}

As already mentioned, the presence of air in a realistic experiment is unavoidable and, even if the pre-existing air is partially removed, there will always be some remnants that will affect the movement of the Satellite. 
In this section we consider the Satellite moving in air and determine the degree of vacuum needed to avoid spoiling the theoretical sensitivity of the proposed experiment. 

\begin{figure*}[t!]
\centering
\includegraphics[angle=0,width=0.47\textwidth]{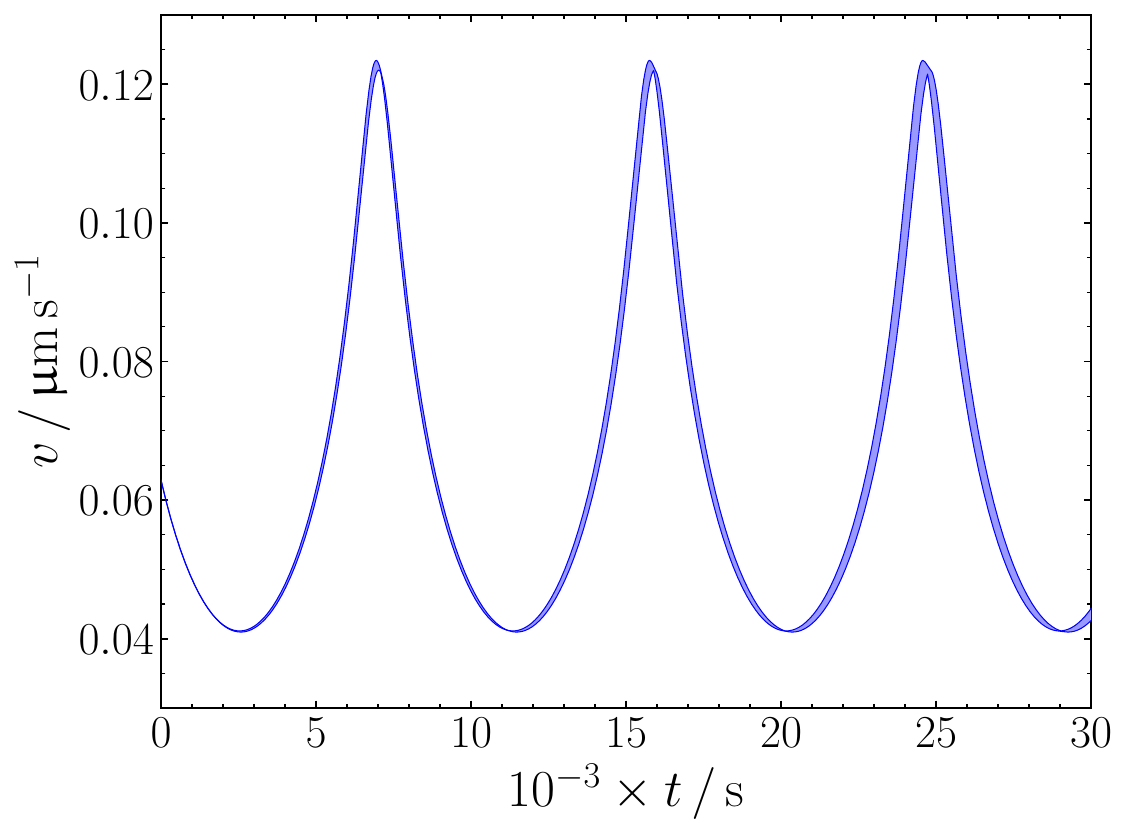} 
\includegraphics[angle=0,width=0.47\textwidth]{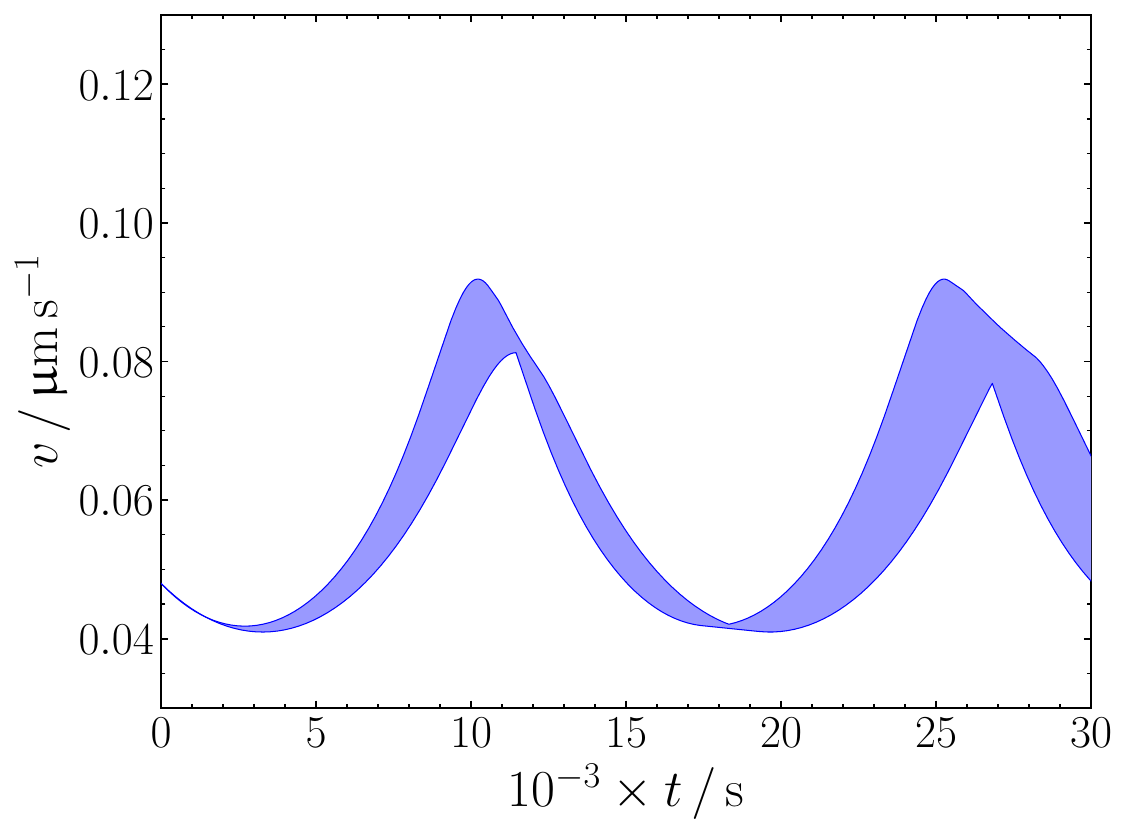} 
\caption{\it 
Absolute velocity $v$ (in $\upmu\textup{m}/\textup{s}$) of the Satellite along its orbit around the Planet as a function of time (only the first few orbits are shown). The shaded regions represent the velocities for all possible values of the nuisance parameters between the Newtonian case ($Q_k = 0$) and the values considered in Fig.~\ref{fig:Limits2}
Left panel: initial conditions corresponding to {\bf Case 1} with maximum nuisance parameters $Q_2 = 10^{-3}; Q_3 = 0.3$ $\upmu\textup{m}$, $Q_4 = 1$ $\upmu\textup{m}^2$; Right panel: initial conditions corresponding to {\bf Case 2}, with maximum nuisance parameters $Q_2 = 10^{-2}; Q_3 = 5$ $\upmu\textup{m}$, $Q_4 = 10$ $\upmu\textup{m}^2$.
}
\label{fig:velocity}
\end{figure*}

To include the effect of air, we first determine if we are in a laminar or a turbulent regime. To do so, we compute the velocity of the Satellite when moving around the Planet. The results for the modified-Newtonian case are shown in Fig.~\ref{fig:velocity} for {\bf Case 1} (left panel) and {\bf Case 2} (right panel). In each panel, we represent the velocity of the Satellite as a function of time for all possible values of the nuisance parameters going from the Newton case  ($Q_k = 0$) 
to the maximum values shown in the panels of Fig.~\ref{fig:Limits2}. 
It can be seen that the absolute velocity of the Satellite, $v_{\rm S}$, is extremely small in both cases, between $0.04$ $\upmu$\textup{m}/s and $0.12$ ($0.09$) $\upmu$\textup{m}/s for {\bf Case 1} ({\bf Case 2}), respectively.
The Reynolds number for a sphere with radius $R_{\rm S} = 5$ $\upmu$\textup{m} 
(the Satellite) moving at this speed in air at ambient temperature, $T = 25$ ${}^\circ$C, is $Re = {\cal O} \left ( 10^{-7} \right)$. 
For a spherical object moving at such low Reynolds number through a Newtonian fluid (such as air) we can safely assume to be in the laminar regime and use the Stokes' law to take into account the effect of air viscosity on the Satellite motion. To include it, eq.~\ref{eq:newtonforce} needs to be modified to include this dissipative effect:
\begin{equation}
\label{eq:newtonforce}
m_{\rm S} \ddot{{\bm{r}}} = -\bm{\nabla}V(r) - 6 \pi \, \eta \, R_{\rm S} \,  \dot {\bf r} \, ,
\end{equation}
where the second term in the right-hand side is the friction term, $\eta$ is the dynamical air viscosity and $\dot {\bf r}$ the velocity of the Satellite (such that  $v=|\dot{\bf r}|$). 
As before, we have neglected the motion of the Planet, which is negligible for the considered masses.

Under standard atmospheric conditions (with temperature in the range $T \in [-20,400]$ $^\circ$C  and sea-level atmospheric pressure), 
the air viscosity depends mostly on the temperature. A useful empirical formula to get $\eta$ as a function of $T$ at $p = 1$ bar is \cite{AirViscosity}:
\begin{equation}
\label{eq:airviscosity}
\eta_{\rm air} (T, p=1\,\text{bar}) = 2.791 \times 10^{-7} \times  T ^{0.7355} \;  {\rm Pa}\,  {\rm s}\,,
\end{equation}
with $T$ in Kelvin. Throughout this work we will additionally make the reasonable assumption that the viscosity depends linearly on the pressure, that is,
\begin{equation}
\label{eq:airviscositypressure}
\eta_{\rm air} (T, p) = \frac{p}{1 \,\text{bar}} \eta_{\rm air} (T, p=1\,\text{bar})\,.
\end{equation}

\begin{figure*}[h!]
\centering
\includegraphics[angle=0,width=0.47\textwidth]{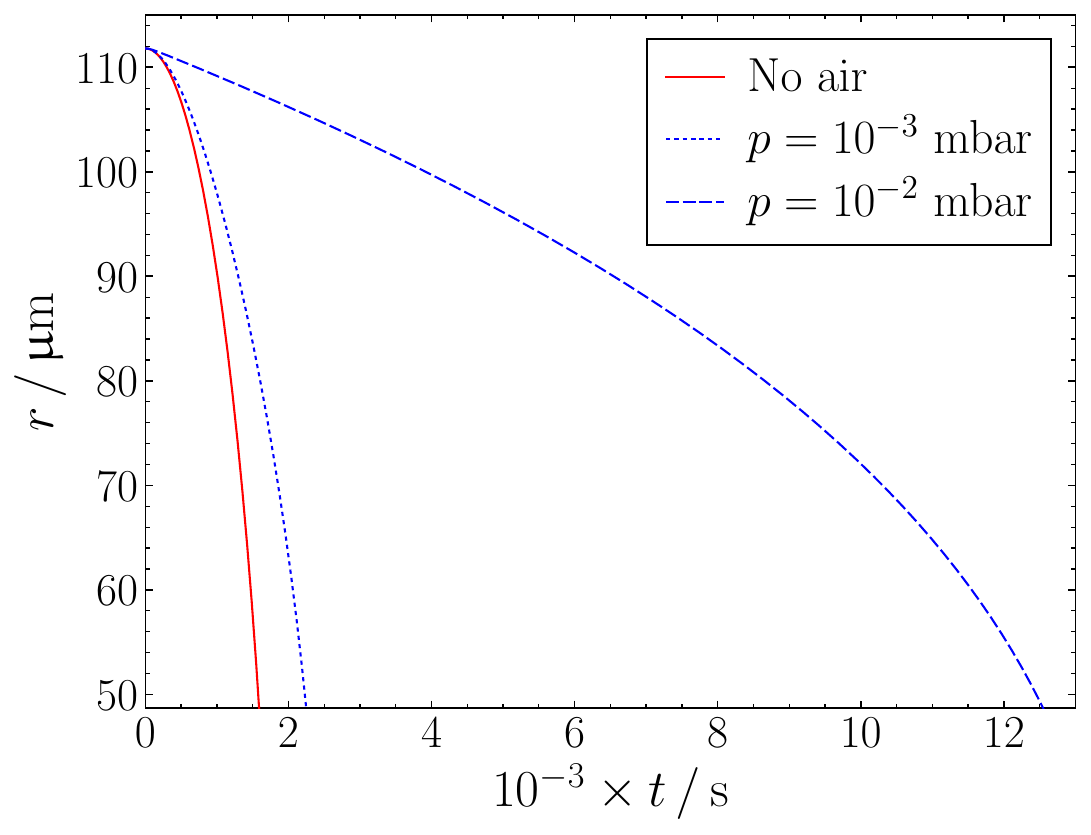} 
\includegraphics[angle=0,width=0.47\textwidth]{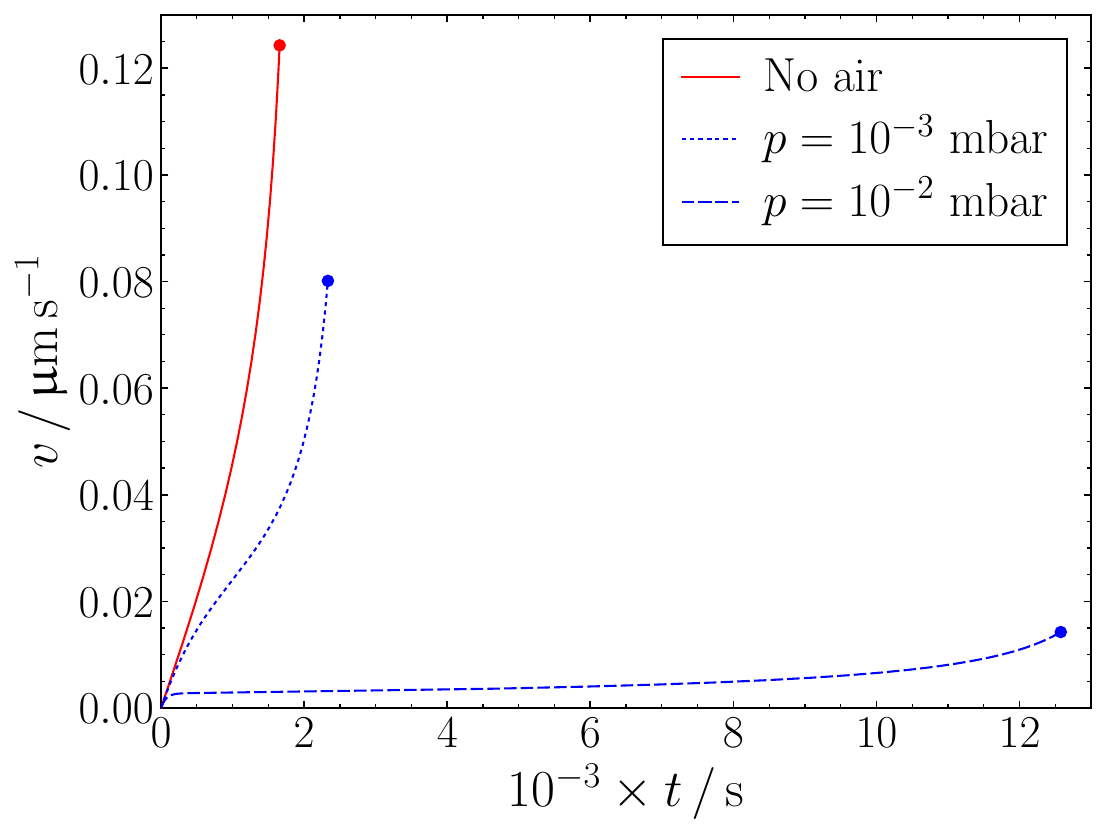} 
\caption{\it 
Position (left panel) and velocity (right panel) of the Satellite moving in a straight line towards the Planet, with initial conditions as follows: $x_0 = 111.8$ $\upmu$\textup{m}, $y_0 = 0; \dot x_0 = 0, \dot y_0 = 0$. No central backgrounds are considered: $Q_k = 0$. The solid red line represents the vacuum case, the dotted blue line an air pressure of $p = 10^{-3}$ mbar and the dashed blue line an air pressure of $p = 10^{-2}$ mbar.
The lines are represented until the Satellite and the Planet collide at
a distance $r_\textup{crash}=R_\textup{S}+R_\textup{P}=48.7\,\upmu\textup{m}$, which is denoted with dots in the right panel.
}
\label{fig:linearviscosity}
\end{figure*}

We will first study the impact of air viscosity on the linear motion of the Satellite approaching the Planet under the effect of the gravitational attraction of the latter, i.e. situations where there is no initial velocity, and the Satellite directly falls onto de planet. This is shown
in Fig.~\ref{fig:linearviscosity} (left panel), where the solid red line
represents the position of the Satellite until it crashes onto the Planet in vacuum, compared to the cases with air pressure $p = 10^{-3}$ mbar (dotted blue line) and $p = 10^{-2}$ mbar (dashed blue line). 
The initial conditions considered are as follows: 
$x_0 = 111.8$ $\upmu$\textup{m}, $y_0 = 0; \dot x_0 = 0, \dot y_0 = 0$, and no central backgrounds have been taken into account: $Q_k = 0$. We can see that, whereas in vacuum the Satellite
will crash onto the Planet in less than 2000 s, 
in the presence of air the time-lapse may increase significantly.
For pressure up to $p = 10^{-3}$ mbar the velocity of the Satellite
increases linearly with time, so that it crashes onto the Planet
in little more than 2000 s. On the other hand, when $p \sim 10^{-2}$ mbar
$v$ increases so slowly that, for the first 5000 s, the
Satellite moves at approximately constant speed. Eventually, after
several hours, the speed starts to increase and the Satellite
collides onto the Planet in something more than 12000 s.
The velocity behavior can be clearly seen in Fig.~\ref{fig:linearviscosity} (right panel). 
It is immediate to extrapolate these plots
to the situation we observe in everyday life: very small objects do not move under the effect of gravitational attraction of nearby bodies, as air pressure is enough to counter-balance it (just have a look at dust particles lighted by a sunbeam 
in a dark room, and see how they move under the effect of air currents and not of gravitation).  
These results, obtained in the case of linear motion, show us that
it is mandatory to remove the air from the Lab volume in order
to detect the gravitational attraction between the Satellite and the Planet. This will be studied in greater detail in the case
of two-dimensional motion in the rest of this Section, in order
to determine which level of air dilution is needed to retain 
the orbit features observed in vacuum.

\begin{figure*}[h!]
\centering
\begin{tabular}{c}
\includegraphics[angle=0,width=0.93\textwidth]{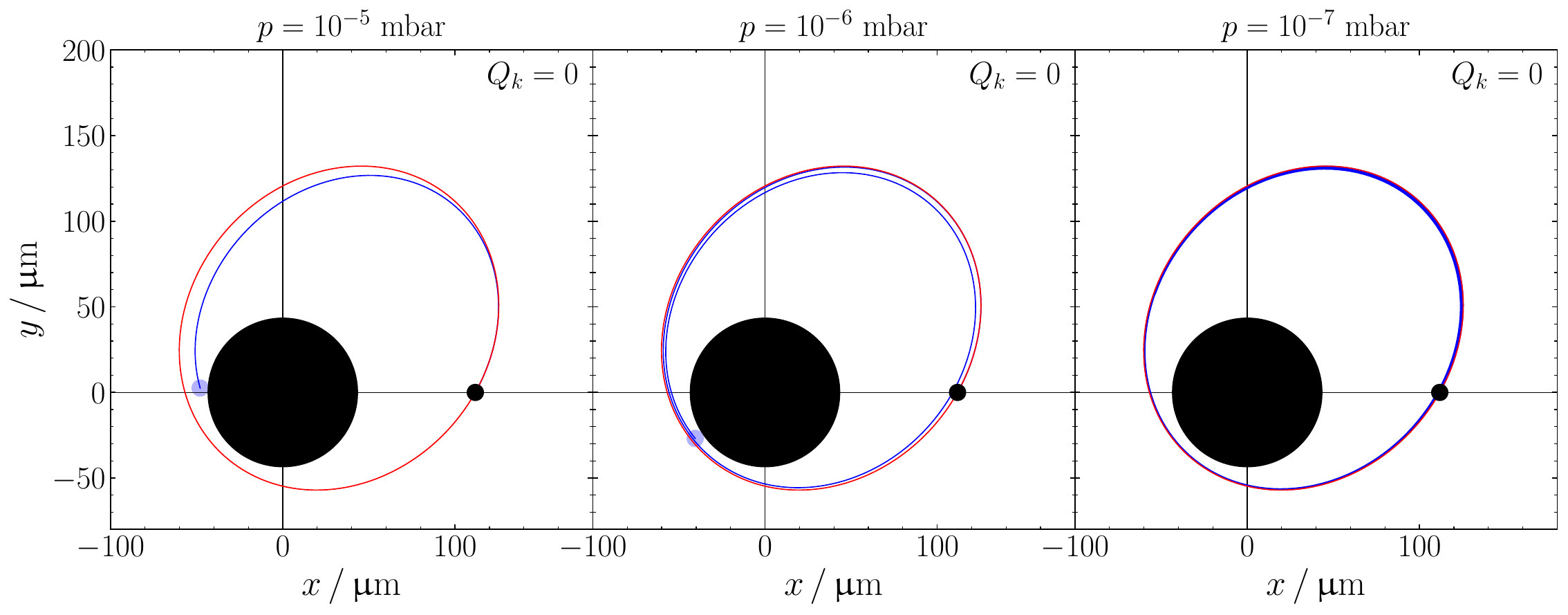}\\
\includegraphics[angle=0,width=0.93\textwidth]{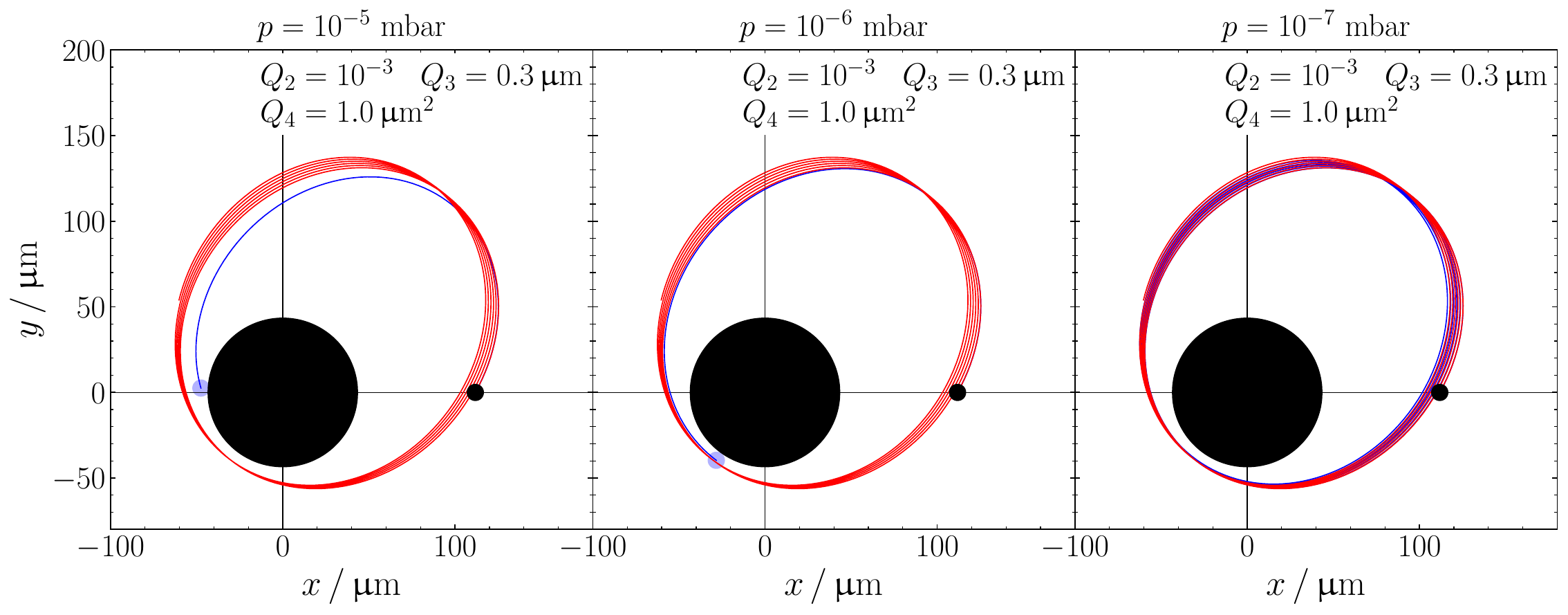}\\
\includegraphics[angle=0,width=0.93\textwidth]{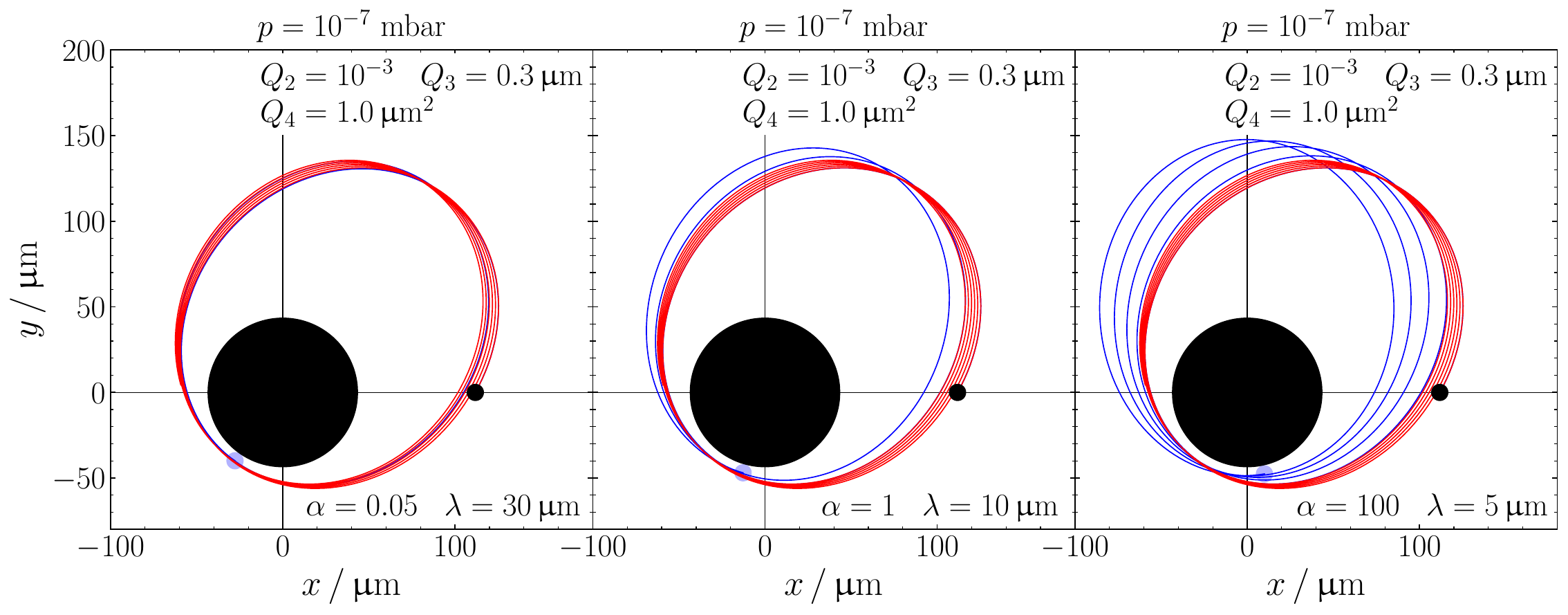}\end{tabular}
\caption{\it 
Bounded orbits of the Satellite around the Planet in the presence of air viscosity (at $T = 25 \,{}^\circ$\textup{C} and air pressure as indicated above each panel) for initial conditions as in {\bf Case 1}. In the top and middle panels, the red (blue) line represents the orbit in the absence (presence) of air viscosity with $Q_k=0$ (top panels) and $Q_ 2 = 10^{-3}$, $Q_3 = 0.3$ $\upmu$\textup{m} and $Q_4 = 1$ $\upmu$\textup{m}$^2$ (middle panels). Bottom panels depict in red (blue) the orbit in the absence (presence) of New Physics, with fixed nuisance parameters $Q_ 2 = 10^{-3}$, $Q_3 = 0.3$ $\upmu$\textup{m} and $Q_4 = 1$ $\upmu$\textup{m}$^2$ and $(\lambda,\alpha)$=$(30\,\upmu\textup{m},0.05)$,  $(10\,\upmu\textup{m},1)$ and $(5\,\upmu\textup{m},100)$, from left to right.
The initial position of the Planet and the Satellite are represented by black dots with $R_{\rm P} = 83.6$ $\upmu$\textup{m} and $R_{\rm S} = 16.4$ $\upmu$\textup{m}, respectively. The orbit is represented for a maximum of 50\,000 s, or until the two spheres collide, in which case the final position of the Satellite is represented in light blue.
}
\label{fig:viscosity1}
\end{figure*}

In Fig.~\ref{fig:viscosity1} (top panels) we show the impact of air viscosity over the trajectory of the Satellite around the Planet for initial conditions chosen as in {\bf Case 1} and no nuisance parameters ($Q_k= 0$). In the left panel, we consider the air viscosity corresponding to $T = 25\,{}^\circ$C with a dilution factor $10^{-8}$ with respect to air density at sea-level atmospheric pressure (i.e., $p = 10^{-5}$ mbar).
Red lines represent the trajectory of the Satellite in vacuum, whereas blue lines are the trajectory in (diluted) air. The Planet and the  Satellite at the initial position are depicted
as two black dots with sizes corresponding to $R_{\rm P} = 43.7$ $\upmu$\textup{m} and $R_{\rm S} = 5$ $\upmu$\textup{m}. The orbit is represented until the two bodies collide (with the Satellite final position depicted in light blue) or up to a maximum of $5\times10^5$ s. 
As $Q_3$ and $Q_4$ are both null, the thick red trajectory corresponds to a stable ellipsis of the Satellite around the Planet, with a Newtonian period $T_{\rm N} \sim$ 2 h 30 min.
We can see that the impact of air viscosity is quite relevant: whereas in vacuum the Satellite kinetic energy is conserved, with $E_\text{K}^\text{per}\sim 8.9\times10^{-21}$ N $\upmu$\textup{m} at the periapsis, in the presence of air it loses energy through friction with the environment and its orbit becomes unstable, crashing into the Planet before completing a single orbit. In order to reduce the impact of air viscosity over the Satellite trajectory, we need to increase the dilution: this is shown in the middle and right panels, corresponding to a pressure $ p = 10^{-6}$ and $p = 10^{-7}$ mbar, respectively. In the latter case, the loss in kinetic energy at the periapsis is just 
$\delta E_K \sim 2.2 \times 10^{-23}$ N $\upmu$\textup{m} per revolution: under these conditions, we can safely study the orbit of S around P for as many revolutions as needed when looking for precession of the orbit. A similar result is obtained when we turn on the nuisance parameters, as represented in Fig.~\ref{fig:viscosity1} (middle panels), where we show again the impact of air viscosity on the orbit of the Satellite for {\bf Case 1} initial conditions, albeit with $Q_2 = 10^{-3}, Q_3 = 0.3$ $\upmu$\textup{m} and $Q_4 = 1$ $\upmu$\textup{m}$^2$. The only significant difference with respect to $Q_k = 0$ is the (small) precession 
of the Satellite orbit in the absence of friction, due to the $Q_3$ nuisance parameter. 

\begin{figure*}[h!]
\centering
\begin{tabular}{c}
\includegraphics[angle=0,width=0.93\textwidth]{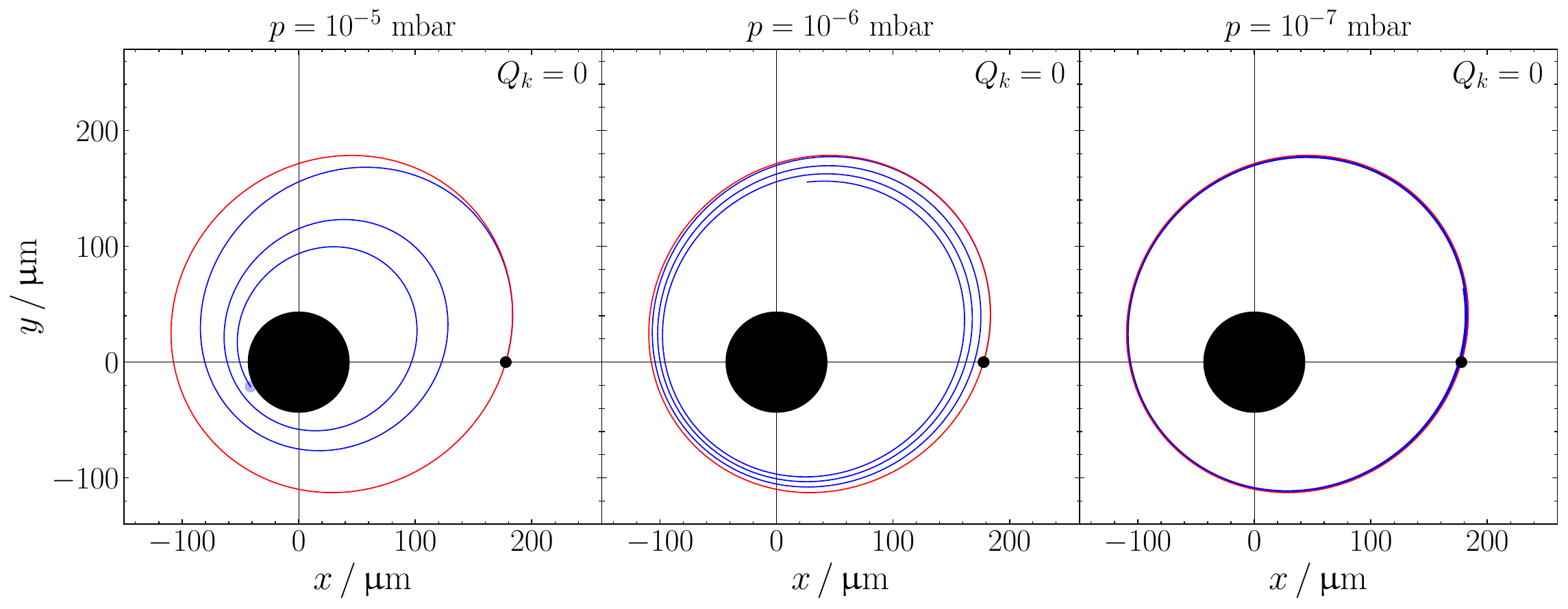}\\
\includegraphics[angle=0,width=0.93\textwidth]{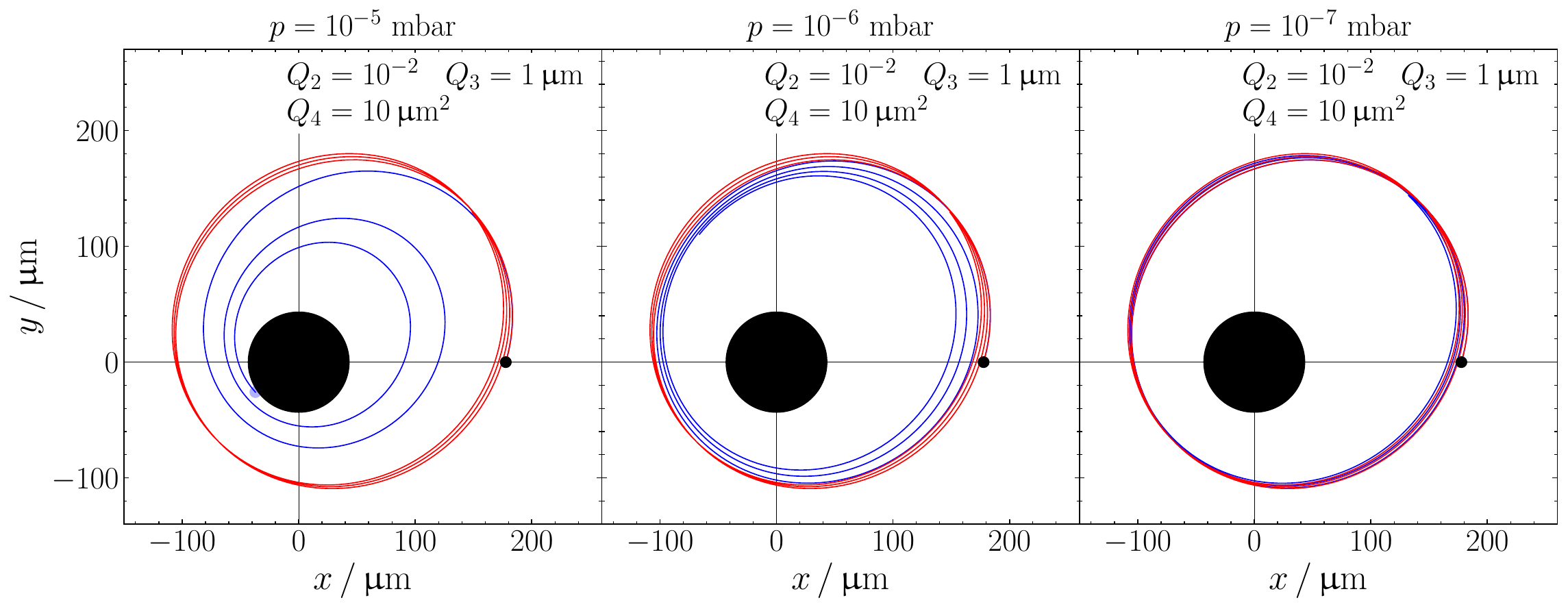}\\
\includegraphics[angle=0,width=0.93\textwidth]{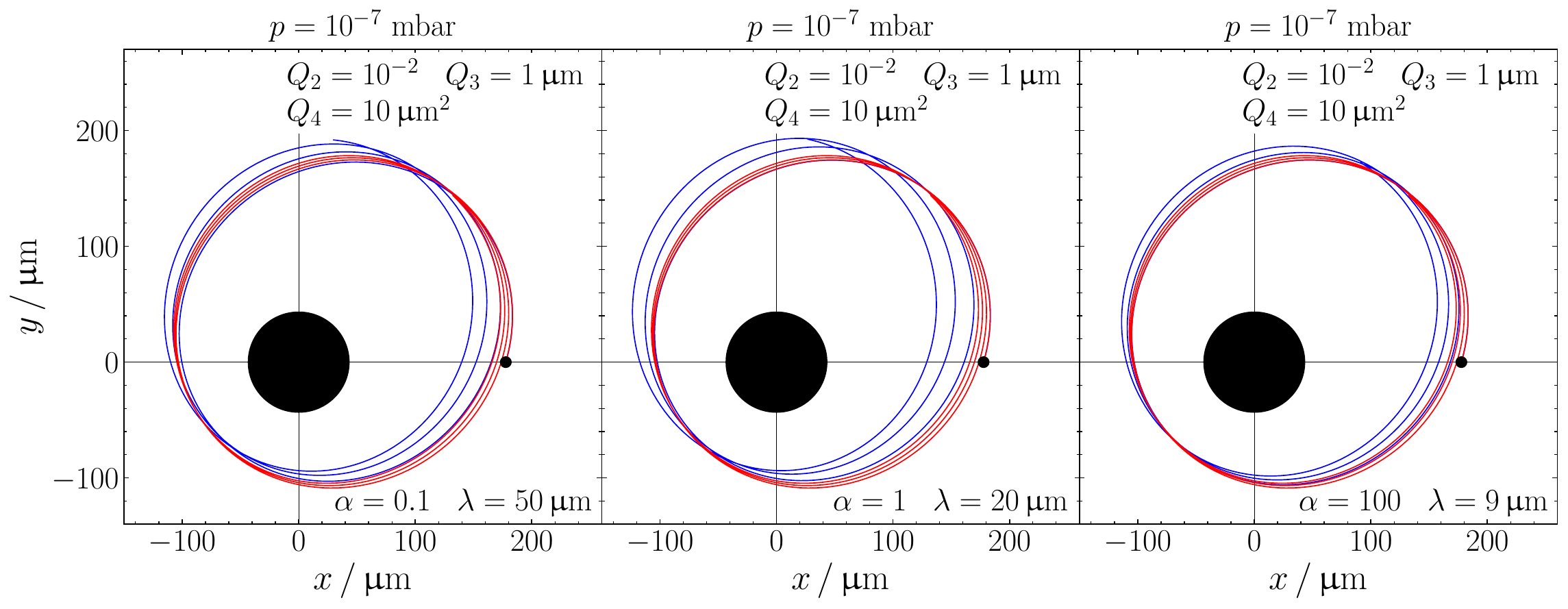}
\end{tabular}
\caption{\it 
Bounded orbits of the Satellite around the Planet in the presence of air viscosity (at $T = 25 \,{}^\circ$\textup{C} and air pressure as indicated above each panel) for initial conditions as in {\bf Case 2}. In the top and middle panels, the red (blue) line represents the orbit in the absence (presence) of air viscosity with $Q_k=0$ (top panels) and $Q_2 = 10^{-2}$, $Q_3 = 1$ $\upmu$\textup{m} and $Q_4 = 10$ $\upmu$\textup{m}$^2$ (middle panels). Bottom panels depict in red (blue) the orbit in the absence (presence) of New Physics, with fixed nuisance parameters $Q_2 = 10^{-2}$, $Q_3 = 1$ $\upmu$\textup{m} and $Q_4 = 10$ $\upmu$\textup{m}$^2$ and $(\lambda,\alpha)$=$(50\,\upmu\textup{m},0.1)$,  $(20\,\upmu\textup{m},1)$ and $(9\,\upmu\textup{m},100)$, from left to right.
The initial position of the Planet and the Satellite are represented by black dots with $R_{\rm P} = 83.6$ $\upmu$\textup{m} and $R_{\rm S} = 16.4$ $\upmu$\textup{m}, respectively. The orbit is represented for a maximum of 50\,000 s, or until the two spheres collide, in which case the final position of the Satellite is represented in light blue.
}
\label{fig:viscosity2}
\end{figure*}

The effect of the air is qualitatively similar when initial conditions as in {\bf Case 2} are considered. This can be seen in Fig.~\ref{fig:viscosity2} (without and with nuisance parameters in top and middle panels, respectively):  to get a negligible impact from air viscosity during the first 50000 s with respect to the trajectory expected in vacuum, we need to dilute the air by a factor $10^{-10}$ with respect to air density at sea-level pressure, {\em i.e.} to achieve a pressure $p = 10^{-7}$ mbar. Notice that, in order to reduce precession due to $Q_i$'s at a minimum, we have considered more stringent bounds on $Q_3$ with respect to those considered in Fig.~\ref{fig:Limits2}. Also note that in this case, even if the air friction produces a non-negligible energy loss, the Satellite is able to complete a number of revolutions around P before colliding, since the initial conditions were much more conservative.

\begin{figure*}[h!]
\centering
\begin{tabular}{c}
\includegraphics[angle=0,width=0.93\textwidth]{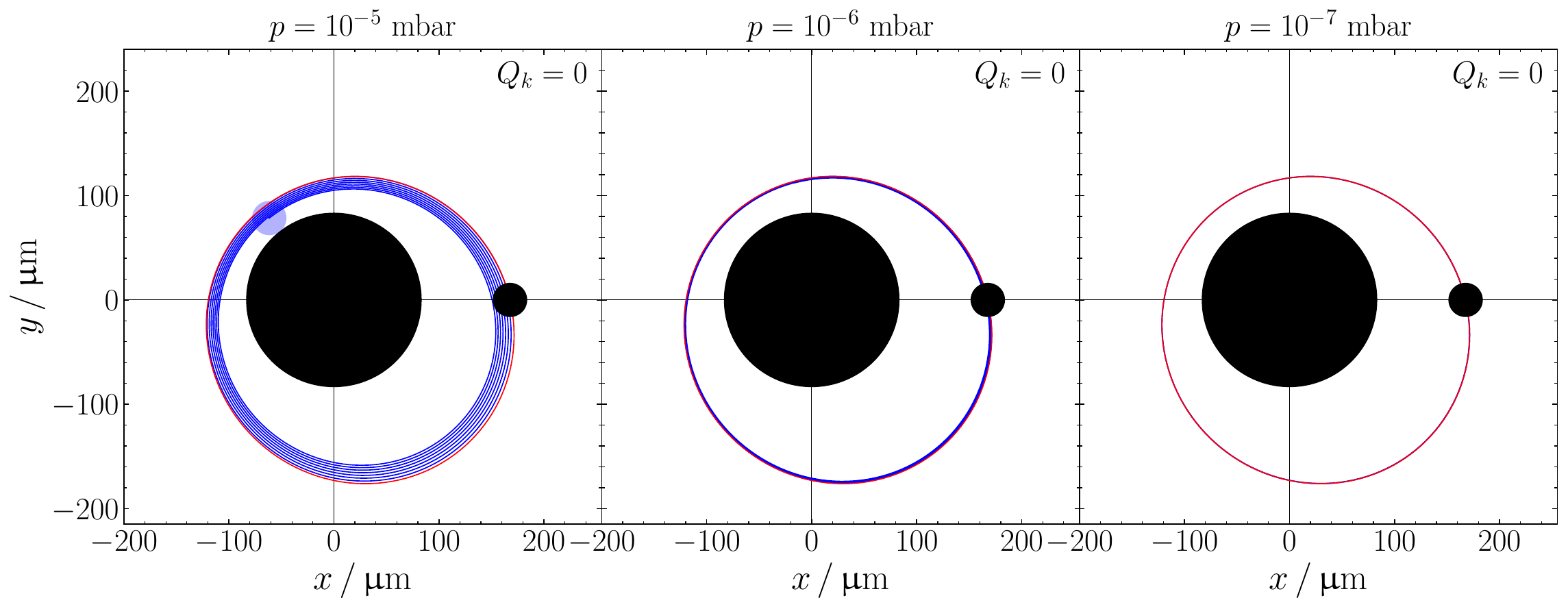}\\
\includegraphics[angle=0,width=0.93\textwidth]{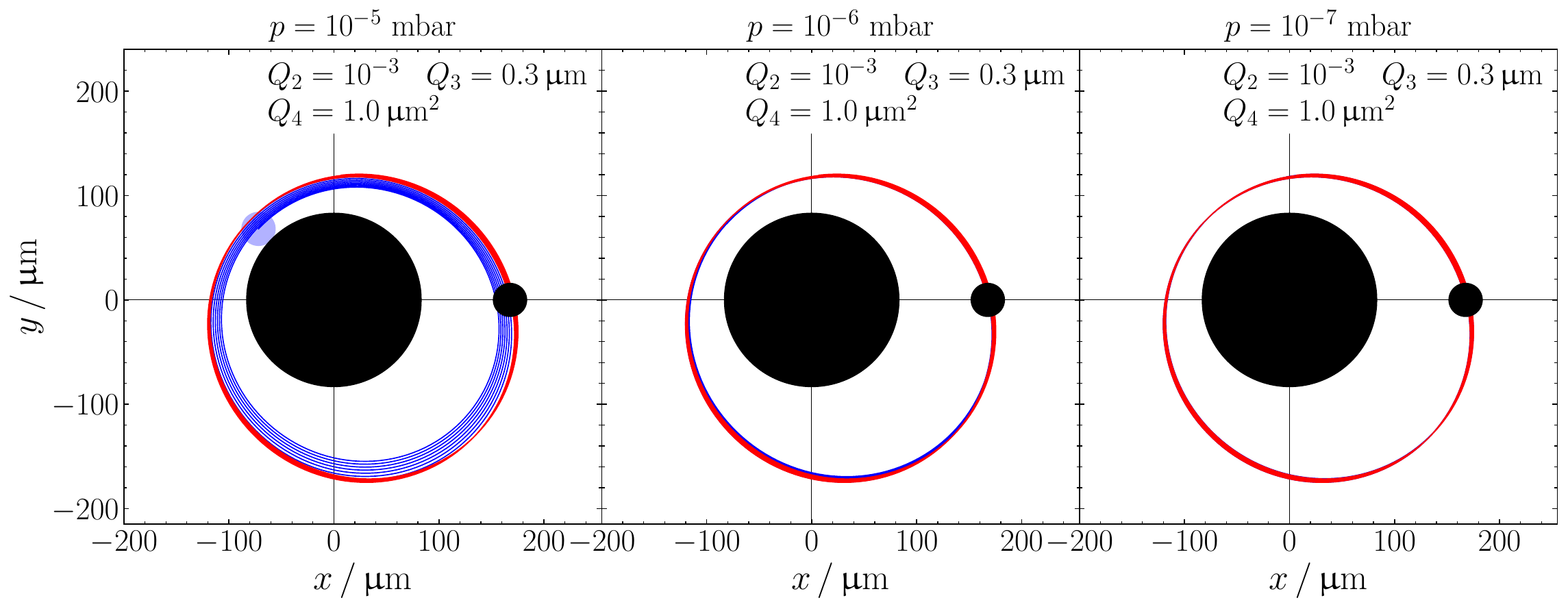}\\
\includegraphics[angle=0,width=0.93\textwidth]{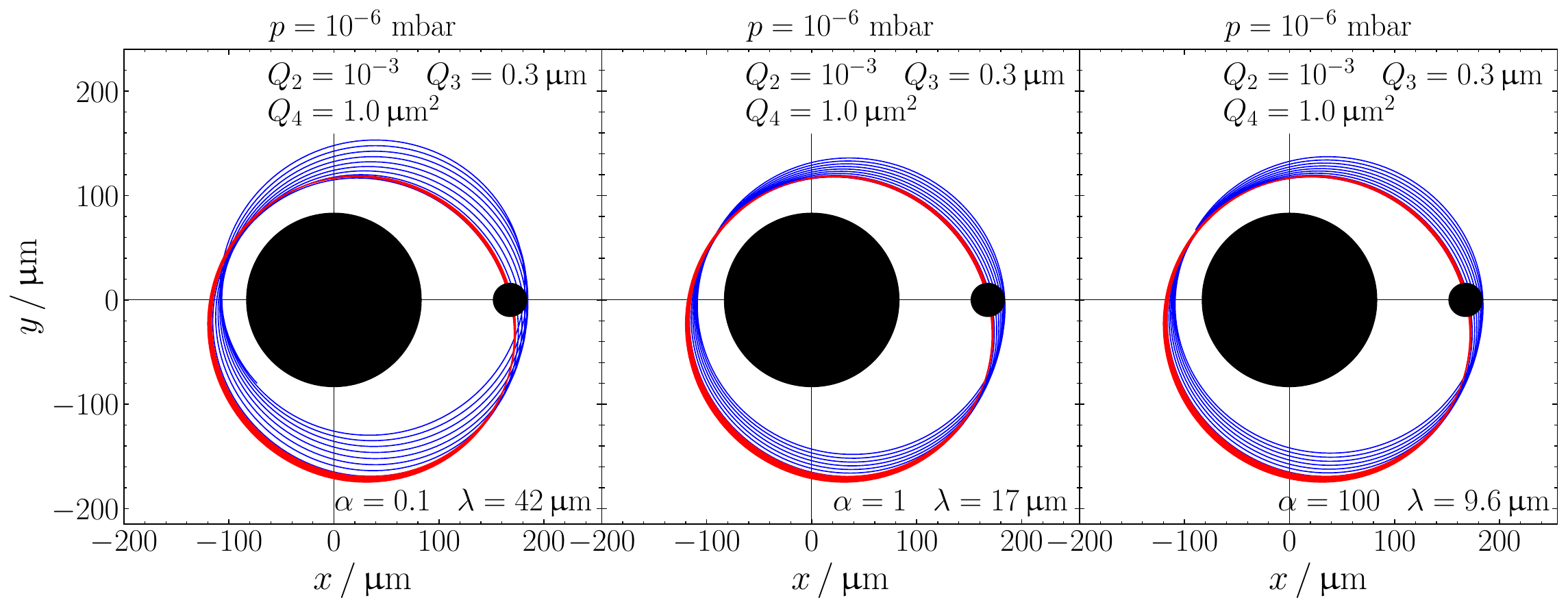}\end{tabular}
\caption{\it 
Bounded orbits of the Satellite around the Planet in the presence of air viscosity (at $T = 25 \,{}^\circ$\textup{C} and air pressure as indicated above each panel) for initial conditions as in {\bf Case 3}. In the top and middle panels, the red (blue) line represents the orbit in the absence (presence) of air viscosity with $Q_k=0$ (top panels) and $Q_2 = 10^{-3}$, $Q_3 = 0.3$ $\upmu$\textup{m} and $Q_4 = 1$ $\upmu$\textup{m}$^2$ (middle panels). Bottom panels depict in red (blue) the orbit in the absence (presence) of New Physics, with fixed nuisance parameters $Q_2 = 10^{-3}$, $Q_3 = 0.3$ $\upmu$\textup{m} and $Q_4 = 1$ $\upmu$\textup{m}$^2$ and $(\lambda,\alpha)$=$(42\,\upmu\textup{m},0.1)$,  $(17\,\upmu\textup{m},1)$ and $(9.6\,\upmu\textup{m},100)$, from left to right.
The initial position of the Planet and the Satellite are represented by black dots with $R_{\rm P} = 83.6$ $\upmu$\textup{m} and $R_{\rm S} = 16.4$ $\upmu$\textup{m}, respectively. The orbit is represented for a maximum of 50\,000 s, or until the two spheres collide, in which case the final position of the Satellite is represented in light blue.
}
\label{fig:viscosity3}
\end{figure*}

Some changes to the experimental setup considered in Refs.~\cite{Donini:2016kgu,Baeza-Ballesteros:2021tha} can be put into place to reduce the impact of air viscosity and, thus, 
soften the need for air dilution in the laboratory volume. The most important optimization we can introduce is to modify the mass of the Satellite: a heavier Satellite will move through air with a smoother flow than a smaller one. If we increase its mass (and its radius), though, the Planet mass has to be increased, too. 
Eventually, the initial conditions must be also modified: the initial distance $r_0$ and the initial angular velocity should be increased, such that collision between the two bodies is avoided and most of the features of the orbit that have been studied
previously may be retained. In Fig.~\ref{fig:viscosity3}, we present our results for a setup in which the mass of the Planet is $M_{\rm P} = 5.25 \times 10^{-5}$ g (with a radius $R_{\rm P} = 83.6$ $\upmu$\textup{m})
and the mass of the Satellite is $m_{\rm S} = 4.2 \times 10^{-8}$ g (with a radius $R_{\rm S} = 16.4$ $\upmu$\textup{m}). The initial conditions are defined as:
\begin{itemize}
\item  {\bf Case 3}: $r_0=167.7$ $\upmu$\textup{m}, $\dot{r}_0=30.6$ nm s${}^{-1}$ and $\dot{\theta}_0=786.2$ $\upmu$rad s${}^{-1}$; 
          for which we have $r_{\rm apo}\sim 187$ $\upmu$\textup{m}, $r_{\rm peri} = 111$ $\upmu$\textup{m} and $T_{\rm N} \sim 1$ h 42 min.
\end{itemize}
Notice that, although the distance between the center of mass of the Planet and the Satellite at the periapsis is $111$ $\upmu$\textup{m}, the distance between their surfaces is just $d = 11$ $\upmu$\textup{m} at the apoapsis.
In top panels we show our results with no nuisance parameters, $Q_i = 0$, whereas in the middle panels we consider $Q_2 = 10^{-3}$, $Q_3 = 0.3$ $\upmu$\textup{m} and $Q_4 = 1$ $\upmu$\textup{m}$^2$, as used in the left top panel of Fig.~\ref{fig:Limits2} . 

The message of the careful analysis we have carried out in this Section is that, in order to look for precession of the orbit of the Satellite around the Planet in a realistic environment, we need a significant amount of dilution of the air inside the laboratoy volume. If using the setup studied in Refs.~\cite{Donini:2016kgu,Baeza-Ballesteros:2021tha}, that was optimized in vacuum, an air pressure $p = 10^{-7}$ mbar is needed both for initial conditions as in {\bf Case 1} and {\bf Case 2}. For {\bf Case 3}, 
that same air pressure would completely eliminate the energy loss due to air friction. Alternatively, one could relax the conditions on the pressure to be  $p = 10^{-6}$ mbar and still study the orbit precession. Once we turn on New Physics, we represent in the bottom panels of Figs.~\ref{fig:viscosity1},~\ref{fig:viscosity2} and~\ref{ fig:viscosity3} the trajectory of the Satellite around the Planet in absence of New Physics but with nuisance parameters (in red) and in presence of New Physics (in blue) for all three cases. Three benchmark points in the $(\lambda, \alpha)$ plane, lying approximately on the ultimate sensitivity line  of each case (see Fig.~\ref{fig:Limits2}), are shown. 
In Fig.~\ref{fig:viscosity1} we represent {\bf Case 1}, 
with $Q_2 = 10^{-3}, Q_3 = 0.3$ $\upmu$\textup{m} and $Q_4= 1$ $\upmu$\textup{m}$^2$, for $(\lambda,\alpha)$ = (30 $\upmu$\textup{m}, 0.05),  (10 $\upmu$\textup{m}, 1) and (5 $\upmu$\textup{m}, 100), from left to right.  Fig.~\ref{fig:viscosity2}  refers to {\bf Case 2}, with $Q_2 = 10^{-2}, Q_3 = 1$ $\upmu$\textup{m} and $Q_4= 10$ $\upmu$\textup{m}$^2$, for ($\lambda, \alpha$) = (50 $\upmu$\textup{m}, 0.1),  (20 $\upmu$\textup{m}, 1) and  
(9 $\upmu$\textup{m}, 100), from left to right.  In all these cases $p=10^{-7}$ mbar is assumed.
Finally, results obtained in the newly optimized {\bf Case 3} are shown
in  Fig.~\ref{fig:viscosity3}, with an air pressure $p = 10^{-6}$ mbar, and $Q_2 = 10^{-3}, Q_3 = 0.3$ $\upmu$\textup{m} and $Q_4= 1$ $\upmu$\textup{m}$^2$, for $(\lambda,\alpha) =$ (42 $\upmu$\textup{m}, 0.05),  (17 $\upmu$\textup{m}, 1) and (9.6 $\upmu$\textup{m}, 100), from left to right. 

In this last case, one observes how the trajectory in presence of New Physics precedes counter-clockwise at a rate of approximately $5^\circ$
per revolution. As we have already mentioned, while $p=10^{-6}$ mbar ensures a negligible effect of the air friction, a slightly higher pressure would also be acceptable, allowing to study the movement of the Satellite around the Planet for several revolutions before the two collide. In any case, once a careful calibration of the experiment is carried out, a better optimization of the setup could be performed and, possibly, air dilution requirements could be further relaxed. 
For the time being, we will consider the setup given in {\bf Case 3}.

\section{The experimental setup}
\label{sec:setup}

\begin{figure}
    \centering
    \includegraphics[width=0.4\textwidth, trim={7cm 1.5cm 0cm 0cm},clip]{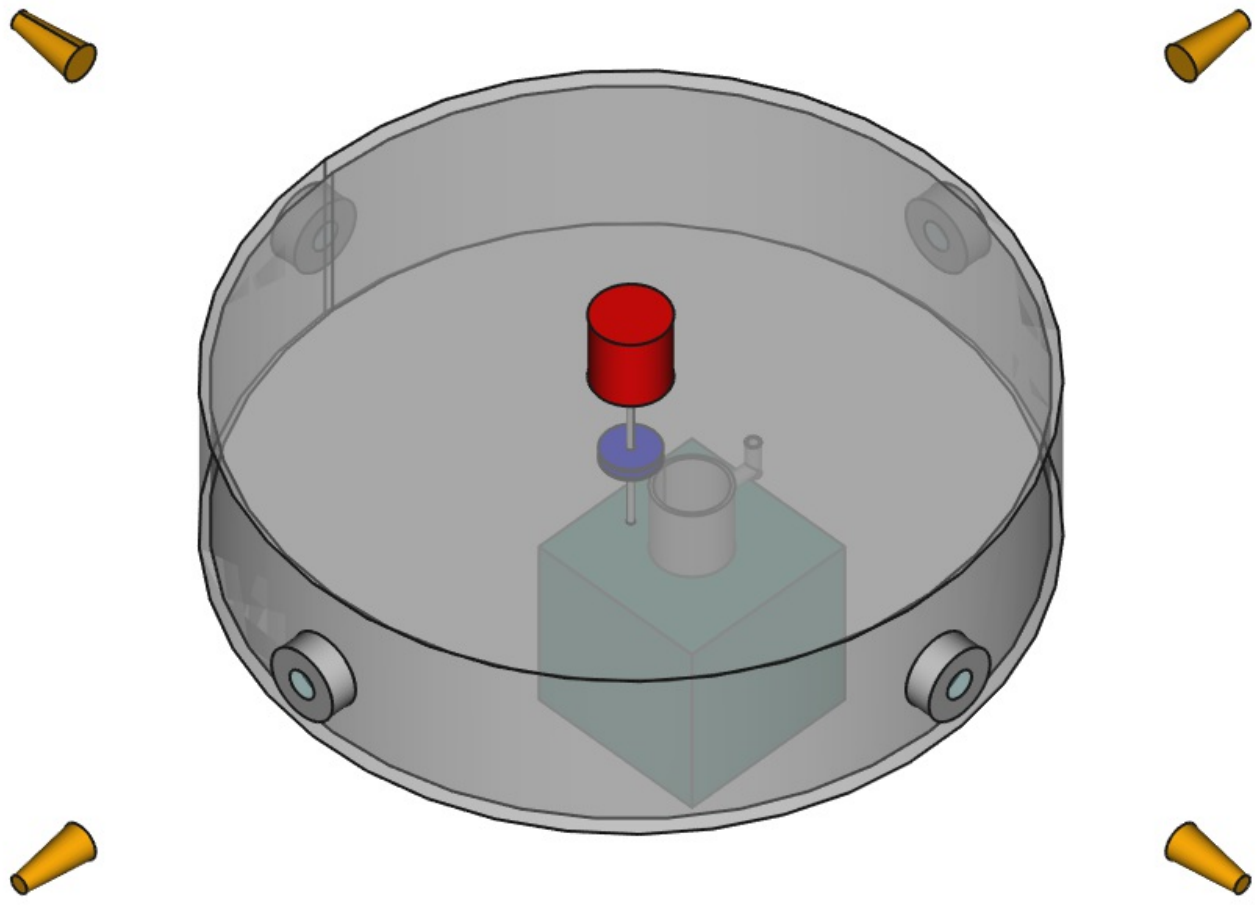}
    \caption{\it Simplified sketch of the experimental proposal with distances not to scale. The magnets, painted blue, are placed on the center of a large vessel. The red cylinder on top represents the manipulator to adjust the magnets distance. Vacuum is done through two ports in the bottom, with the pumping system represented with a green cube. The orbit is recorded by four CMOS cameras (orange cones) outside of the vessel, through feedthroughs equipped with lenses. Additional feedthroughs can be added between the optical ones to accommodate for the injection system.}
    \label{fig:exp_setup}
\end{figure}

As we have shown in the previous Section, it is mandatory to modify
the setup proposed in Refs.~\cite{Donini:2016kgu,Baeza-Ballesteros:2021tha} in order to attain a sensitivity close to that expected
in the ideal case with no air friction. In this Section we will detail the elements that a realistic experimental set-up should include, so that the results in this work could be actually tested. As in any real experiment, we expect that many of the characteristic elements given here will be modified after a calibration of the setup allows to find better alternatives. However, we will discuss the features of a schematic representation of the ideal setup, which is represented in Fig.~\ref{fig:exp_setup} (distances not to scale).


\subsection{The Planet and the Satellite}

We will call in the following as the Lab the volume between 
two magnets used to levitate the Satellite. The Planet will be inserted into the Lab before air removal. Once we achieve the desired
vacuum level, the Satellite will be introduced in the Lab. Eventually,
we will put it into motion around the Planet with photo-irradiation. 

The optimal choice for the material of which the Planet is made
is one with the highest density at room temperature
and under conditions of very low pressure. This way, we may have a Planet with the smallest radius possible whilst having the desired mass, giving us more flexibility when choosing the initial conditions. Platinum is the optimal choice, thanks to its very high density at room temperature and pressure, whereas gold would be the simplest alternative. A key feature to choose the optimal material for the Planet should be the simplicity to shape it into a (reasonably) spherical object with the right dimensions and mass. Also, it should be easy to manipulate to  insert it into the Lab and kept it stationary. For the time being, we will consider a platinum sphere with radius $R_{\rm P} = 83.6$ $\upmu$\textup{m} and mass $m_\text{P} = 5.25 \times 10^{-5}$ g, as discussed in the previous section, located at the center of the setup in a fixed position. As explained earlier, these parameters are slightly different from 
those in Refs.~\cite{Donini:2016kgu,Baeza-Ballesteros:2021tha}, as
we are taking into account the presence of some residual air into
the Lab. 

The requirements for the Satellite are completely different than
those given for the Planet. In this case, the key feature is that we
need a diamagnetic material, as we need the Satellite to levitate in a magnetic field in order to cancel the effect of the Earth's gravitational field. As in Refs.~\cite{Donini:2016kgu,Baeza-Ballesteros:2021tha}, the Satellite is a pyrolytic graphite sphere with radius $R_{\rm S} = 16.4$ $\upmu$\textup{m} and mass $m_\text{S} = 4.2 \times 10^{-8}$ g. Pyrolytic graphite has a diamagnetic susceptibility $\chi = - 4 \times 10^{-4}$  \cite{Simon:2000} and allows for magnetic
levitation. The mass of the Satellite is approximately 20 times larger
than in Refs.~\cite{Donini:2016kgu,Baeza-Ballesteros:2021tha}, again 
to reduce the effect of air friction. 

The materials used should allow to fabricate both bodies with good finishing, preventing imperfections that may affect the symmetry of the system and, eventually, the measurements themselves.\footnote{The impact of asphericity of the Planet or of the Satellite on the latter orbit has not been studied. It is supposed to be negligible, given the average distance of the Satellite from the Planet at each point of its trajectory, but a careful study of these effects could be theoretically computed and tested on when the real experiment will be realized.} While we are unaware of commercial options for micrometer scale pyrolytic graphite spheres, spheres of even smaller size, of the order of hundred nanometers, of such material have been already been formed by pyrolysis of propylene \cite{LI2017428}. Larger size sphere acquisition will be further investigated. If obtaining such a sphere made of pyrolytic graphite is unfeasible, a number of alternatives in the few tens of micrometer scale may be useful: commercial plastic spheres from different materials such as polyethylene are available at low cost \cite{cospheric,bioparticles}, as well as regular graphite spheroids formed by heat treatment of medium carbon steel \cite{LI2017428} or bismuth spheres produced by thermal alcohol reduction \cite{HUANG2022133239}. Of course, given the differences in density and magnetic susceptibility of each material, the magnetic field would need to be optimized and adapted in case the pyrolytic graphite can not be used.

\subsection{The Levitation System}

\begin{figure}
    \centering
    \includegraphics[width=0.5\textwidth, trim={0cm 0cm 0cm 0cm},clip]{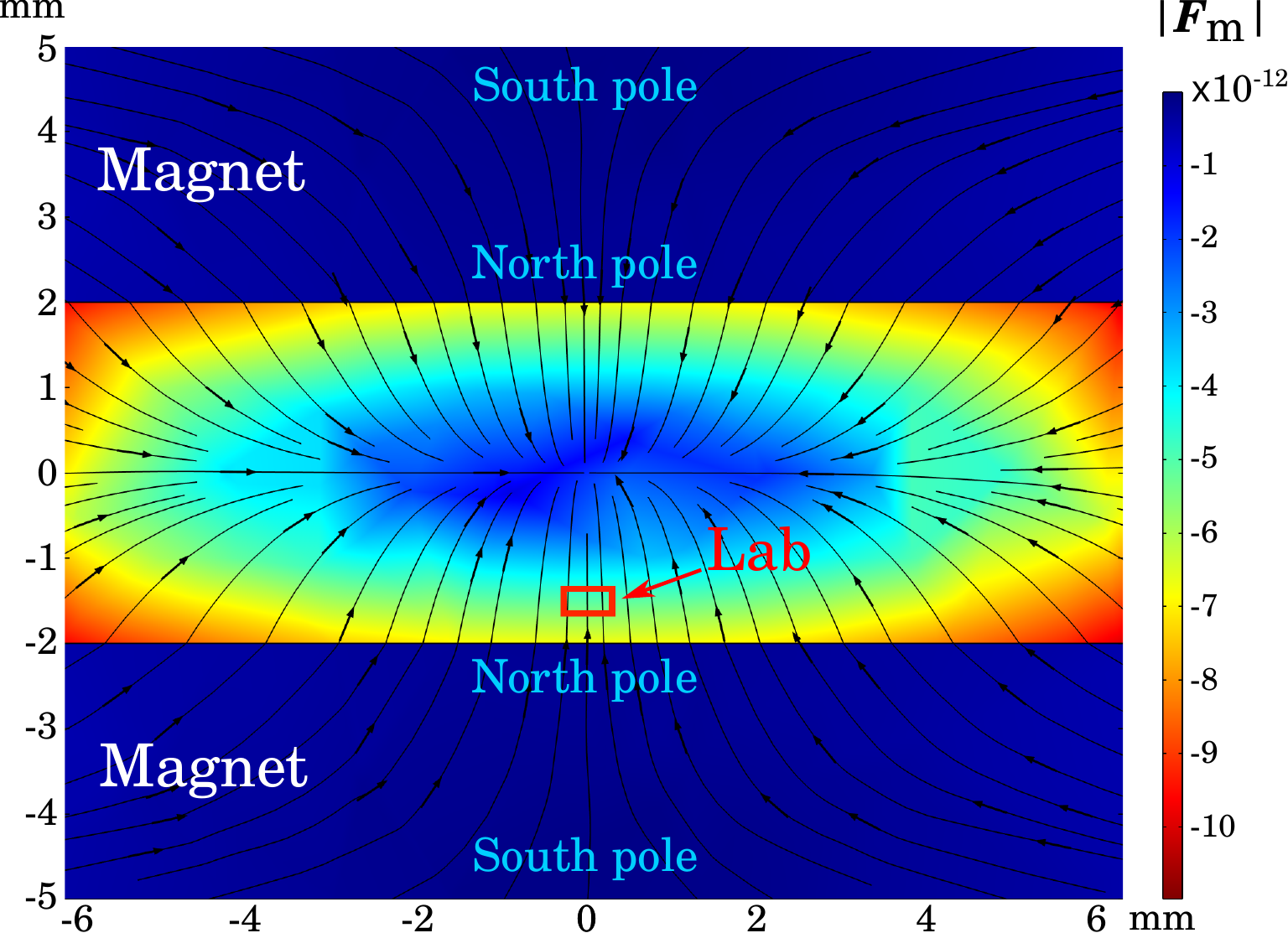}
    \caption{\it The magnetic force field of the considered setup on the Satellite for two $1.25$ \textup{T} neodymium magnets, in units of $10^{-10}$ \textup{N}. The separation between the two magnets (depicted in dark blue) is $4$ \textup{mm}, their radius is $15$ \textup{mm} and they are arranged with the north faces facing each other. The equilibrium position of the Satellite is approximately $300$ $\upmu$\textup{m} above the north pole of the lower magnet (in the vertical legend it corresponds to the yellow region). The Lab zone is represented by a red rectangle. We used the COMSOL Multiphysics® software to perform the modeling \cite{comsol,multiphysics}.
    }
    \label{fig:magneticfield}
\end{figure}

The Satellite needs to move freely around the Planet and, therefore, we need to somehow cancel Earth's gravitational field. The simplest way for doing this for the Satellite mass range considered here is the use of a relatively strong magnetic field ($\sim$ 1 T) in the Lab volume, as already introduced in Ref.~\cite{Baeza-Ballesteros:2021tha}. This can be achieved with either electromagnets or permanent neodymium magnets. The design of the magnetic field should allow for a sufficiently homogeneous field in the plane of the orbit that provides stable conditions for the Satellite to orbit around the Planet, without introducing a drift on the orbit plane\footnote{Note that inhomogeneities in the magnetic field, due to the symmetry of the setup, are expected to generate a quadratic potential to first order, and so will not affect the precession of the Satellite's orbit}. In Fig.~\ref{fig:magneticfield} we present the force field acting on the
pyrolytic graphite Satellite,
produced by two commercial cylindrical Neodymium magnets (depicted in dark blue) with a magnetic field $|B| = 1.25$ T, whose north faces are facing each other. The separation between the two magnets is 4 mm. The radius of the two magnets is 15 mm. The red rectangle just above
the north face of the lower magnets represents the region in which the
Satellite is expected to orbit around the Planet. In this region the
gravitational force of the Earth on the Satellite is canceled by the upward pointing
dipolar magnetic force. As it can be seen by the arrows on the field lines, 
the magnetic force points upwards in the lower part of the volume between the two magnets and downwards
in the upper part, with null vertical field midway between the
two. This particular configuration of the force field is
a consequence of the negative magnetic susceptibility of the pyrolytic
graphite. If we were to increase the Satellite mass, the red rectangle would move downwards. In order to keep it sufficiently
far from the upper face of the lower magnet (so as to avoid boundary effects), we should increase the magnetic field to compensate for the
increase in the mass. 

The upper magnet will be coupled to a vertical manipulator to adjust the magnetic field by modifying the distance between poles. This will allow to adjust possible misalignments between the Planet and the Satellite in case the equilibrium position of the latter differs from what is initially expected, due to tolerances of the fabrication process which may end up in a different density than desired. It is important to note that commercial manipulators with resolution in the $\upmu$\textup{m} scale are already available \cite{manipulator_pfeiffer}, something that enables precise fine-tuning of the field. The manipulator will be coupled to a port aligner to adjust the tilt of the upper magnet and ensure both magnets are completely parallel.
The impact of any difference between the average mass of the upper and lower magnets, that would result in a correction to the (vertical) Earth gravitational field, can be removed by adjusting the upper magnet position.

Notice that an alternative possibility to levitate the Satellite would be using optical levitation. In this case, the Satellite should be made of a dielectric material. This possibility will not be studied here, where we will stick to the original proposal of magnetic levitation, but will be explored in a future work.

\subsection{The Vacuum System}

As it was shown in Sect.~\ref{sec:viscosity}, we need to operate our setup in a vessel with good vacuum conditions (at the level of $10^{-6}$ mbar or better). With this in mind, it is desirable to minimize the number of elements inside the chamber as well and use low-outgassing materials.

The experiment would be housed in an electro-polished stainless steel to reduce outgassing to a minimum. The magnet system will be located inside, in the middle of the chamber, as depicted in Fig.~\ref{fig:exp_setup}. The vessel shall have 4 lateral ports aligned with the magnet system, distributed at 90$^\circ$ angles. Each of these ports would end in a CF viewport coupled with a lens. An additional port shall host the system to put the Satellite in orbit.

A turbo pump in continuous operation will be used to achieve and keep the target vacuum level while performing the experiment. Given the simplicity of the system and the reduced number of elements inside the vessel, the vacuum at the system should be considerably better than the minimum $10^{-6}$ mbar goal. The turbo pump will be connected to the vessel through a guillotine valve. An auxiliary connection will connect the pump to the vessel through a 1/4 inch VCR connection. A primary scroll pump connected to the turbo pump completes the vacuum system.

Vacuum shall be done in two stages. First of all the Planet will be
introduced in the vessel, coupling it rigidly with it. Then, 
the scroll pump will pump down the vessel through the VCR connection, while the guillotine valve remains closed. The small cross-section of the connection will substantially reduce the suction force, but an additional flow controller may be needed in this connection to ensure the Planet is not displaced during the first vacuum stage. Once a vacuum of $10^{-2}$ mbar is reached, the guillotine valve will be opened and the turbo pump will be turned on. Such system should reach the target vacuum level in a couple of hours. Vibrations of the Planet around the center of the vessel should be measured after this stage, in order to introduce them in the theoretical expectation for the Newtonian orbit. Eventually, the Satellite can be injected into orbit and the experiment can be run.


\subsection{The Satellite Injection System}

One of the most important requirements to perform the experiment proposed here is the capability to put the Satellite into a given orbit 
around the Planet with reasonably good precision. This means
that we need a good control on how to fix the initial conditions and, 
in particular, on the way used to put the Satellite into motion 
with given angular and radial velocities. 
We need to do that inside a vacuum chamber and use a system that can be 
decoupled from the planetary system after transferring the initial impulse, so that it will not modify the gravitational field of the setup. Our first choice is to use photo-irradiation \cite{Kobayashi:2012}, but we could also consider different techniques such as, for example, momentum transfer by a third sphere (like in a billiard table). An alternative possibility could be to consider the use of Type-II superconductors to trap the Satellite in a vortex to set the initial momentum mechanically, and then vary the temperature of this superconductor to release the Satellite. 
In this paper we will, however, stick to the first option and describe how to fix the Satellite orbit using photo-irradiation. In order to do so, using a pulsed laser to control the initial conditions should be feasible. As an order of magnitude, a continuous wave laser of around 10 mW of power will exert a force in the order of pico-Newtons, i.e. $F\sim10^{-12}$ N. This order of magnitude is computed for perfectly reflecting spheres \cite{Kim:06}. 
Although this regime could be achieved by coating our Satellite with a suitable material, in fact the order of magnitude still holds for particles with dielectric properties such as those of pyrolytic graphite. By properly controlling the temporal duration of the laser pulse, one can accurately control the momentum transmitted to the Satellite and the initial velocity. Both the magnitude and direction of the Satellite's velocity can be controlled via the laser pulse duration and its propagation direction, respectively. For  $m_\text{S} \sim  10^{-8}$ g and a 1 mW laser with a pulse duration in the microsecond range ($\tau \sim 10^{-6}$ s), one can reach the desired order of magnitude of the initial conditions of $v_i = F \tau /m_\text{S} \sim 10^{-8}$ m/s $= 10^{-2}$ $\upmu$\textup{m}/s. This shows that we have a wide range of parameters of the input laser pulse in order to accurately control the initial velocity of the Satellite, a procedure that will have to be carefully calibrated before starting the measurement of the orbits.

 A central point which is worth reminding is that an error in fixing the initial conditions does not affect whether the Satellite precedes or not around the Planet, as it would only modify the Keplerian period of the orbit. Therefore, if we were able to detect 
 precession, any such error would be irrelevant. However, these uncertainties do play a role when trying to optimize the setup or to determine its sensitivity, due to different reasons. First, errors in the choice of initial conditions eventually imply a wrong prediction of the Newtonian period. In App.~A of Ref.~\cite{Baeza-Ballesteros:2021tha} we study in detail the impact of errors in the initial conditions. We showed that a 1\% error on the initial positions or on the initial velocity implies an error of up to 5\% on the Newtonian period. In addition to this, a wrong choice of the initial conditions
 may also change the Satellite trajectory from a closes orbit to a collisional motion.  We found that errors in the setting of initial conditions below 10\% should not modify significantly the sensitivity of the setup. 



\subsection{Imaging}\label{sec:camera}

In order to reconstruct the orbit of the Satellite around the Planet we need to take images of the system at different times. Due to the long duration of the periods of the system, we consider 1 Hz image rate more than enough 
(remind that, for the range of initial conditions considered here, a typical orbit of the Satellite around the Planet would take a couple of hours).

A system of 4 (symmetrically-located) lateral cameras, one at each side of the system (see Fig.~\ref{fig:exp_setup}), can be used to record the orbit of the satellite. This would allow to determine the three-dimensional path of the Satellite and take into consideration any misalignment between the Satellite and the Planet. The simpler configuration would be to put the cameras outside the vessel, which would simplify the design of the chamber and, at the same time minimize the number of elements inside of it, which is desirable to minimize outgassing and reach the target vacuum level. The symmetric arrangement of the cameras will also reduce the gravitational impact on the system. 

The cameras shall be located in front of the 4 vessel viewports. A possible configuration would be to have the lenses 200 mm away from the planetary system, with the cameras 400 mm away for a 2$\times$ magnification. 16 mm diameter lenses with a focal length of 133.33 mm would produce a depth of field of 0.19 mm, with a circle of confusion of 15 $\upmu$\textup{m}. That depth of field would cover at least half of the depth of the total orbit, and the full depth would be covered with the use of two fronted cameras. While the circle of confusion may seem considerably large at first, its impact would be minimal to the system as the orbit analysis relies on a good knowledge of the position of the Satellite center-of-mass, not on the good definition of the edges. However, when the Satellite is in the periapsis ({\em i.e.} at the minimal distance from the Planet), separating the Planet and the Satellite may not be evident and could have an impact on the Satellite center-of-mass determination. 

For the proposed setup, and using a 6.5 $\upmu$\textup{m} pixel CMOS camera such as Hamamatsu C14440-20UP ($\sim$ 1.2 kg per camera), the image pixel size would be of 3.25 $\upmu$\textup{m} and the center of the Planet could be measured with a precision as low as 0.3 $\upmu$\textup{m} when performing a simple weighted average of the CMOS pixels at low light levels ($\sim$ 10 photons per pixel). While this estimate was obtained with a simple toy model, with an ideal optical system and not including noise from the CMOS, it allows us to conclude that the experiment is feasible, as the extra noise could be compensated by a higher illumination signal, or operating the camera in different acquisition modes. This would allow for increasing the number of frames per second possibly at the cost of increasing the noise per image. For the lowest noise configuration and an image size of 2304$\times$128 pixels, up to 96 frames per second can be acquired, while our goal is just to acquire a single image per second. While it is desirable to keep the acquisition rate to the 1 Hz target, to minimize the volume of data ($\sim$0.5 MB per frame), a higher number of frames can be acquired and later combined offline into a single frame in order to improve the position estimate. Taking all into account and on a conservative scenario, we envision the proposed setup to be able to determine the orbit path with a precision of 
$\sigma \sim$ 1 $\upmu$\textup{m} while keeping the minimum distance to differentiate Planet and Satellite to 30 $\upmu$\textup{m}. 

Note that these numbers could be improved by tuning the various parameters in play. For example reducing the image pixel size, either with a camera of better granularity or a larger magnification, would improve the orbit reconstruction precision. On the other hand, lenses with a smaller diameter would minimize the circle of confusion and, ultimately, shorten the minimum detectable distance between Planet and Satellite. In any case, the performance of the proposed optical setup can be tested and optimized in a dedicated setup prior to building the final experiment.

The position of the four cameras should be carefully measured, in order
to place them symmetrically with respect to the Lab area. In this way, 
we can minimize the effect of their gravitational field on the micro-planetary system. We have computed their impact, assuming 4 ``cameras" of 1.2 kg
each (represented by spherical mass distributions), 
located at a distance of 60 cm from the Lab area. While a single camera would produce a gravitational force on the Satellite of the same order as the Planet, the symmetric configuration lead to a force that is various orders of magnitude lower, and so its impact on the Satellite motion is
negligible. The effects of small asymmetries as well as of the presence of other macroscopic masses close to the experiment will be studied in a future work.

\subsection{Quenching system}

As we have explained above, our microscopic planetary system is composed
of a Planet, considered to be fixed at the center of the system, and a Satellite, which is moving on a bounded (but no necessarily closed) orbit around the former. As a first approximation, we have considered the motion of the Planet due to gravitational attraction of the Satellite negligible. Even within this approximation, however, it is possible that the specific way in which 
the Planet has been introduced into the Lab and rigidly 
fixed in it, or the air pumping procedure, may induce an oscillatory motion of the Planet around the origin of coordinates. 
A comprehensive study of the impact of Planet vibrations on the motion of the Satellite will be presented elsewhere, and in this paper we summarize the results of a preliminary scan of this effect. 
We found the following: 
\begin{enumerate}
    \item Vibrations of the Planet along the vertical axis with small amplitude have a negligible effect on the existence (or not) of precession in the horizontal orbit of the Satellite.
    \item ``Fast" vibrations of the Planet onto the horizontal plane with amplitude smaller than the range of the Planet size do not affect the existence (or not) of precession in the horizontal orbit of the Satellite, although may lead to collision between the two bodies.
    \item The horizontal orbit of the Satellite is modified by ``slow" and ``large" horizontal Planet vibrations.
\end{enumerate}
The precise definition of ``fast" and ``slow" frequency of the horizontal Planet vibrations can be obtained after a careful study of the equations of motion that include (a) the gravitational effect of the Satellite onto the Planet and (b) the motion of the Planet induced by the vibrations. In our preliminary study we have found that only in the case in which the Planet moves with a speed similar to that of the Satellite (whose orbit around the Planet takes a couple of hours to be completed), and gets periodically close to the Satellite, it may alter the its orbit. In that case, resonances may occur and the orbit can be significantly modified. For the time being, we would require that vibrations in the horizontal plane are limited in amplitude to avoid collision of the Satellite and the Planet, with a frequency higher than $\sim 10$ oscillations per hour.

In order to cut ``slow" vibrations, the mechanism with which the Planet is fixed to the Laboratory should be sufficiently ``rigid" (as only ``fast" and ``small" vibrations would be allowed in that case). Once the preparation of the two spheres in our system is specified, it must
include then the possibility of rigid anchoring of the Planet into the Lab.

\section{Sensitivity of a ``realistic" setup}
\label{sec:sens}

We will now perform an updated study of the sensitivity of our setup, 
after taking into account some of the novelties introduced in the previous
Sections with respect to Refs.~\cite{Donini:2016kgu,Baeza-Ballesteros:2021tha}. We will consider the new optimization 
of the initial conditions to minimize the impact of residual air in the lab area and we will take advantage of the new data acquisition
system, consisting of four Hamamatsu C14440-20UP cameras that may provide
the position of the Satellite along its trajectory in real time. For this study, we assume that the nuisance parameters $Q_k$ have been measured
in a calibration phase, and will be considered fixed to the values considered
in Fig.~\ref{fig:Limits2}. The impact of the air pressure will be also
taken into account, fixing the level of air in the
Lab volume. More concretely, we will consider $p\gtrsim10^-5$ mbar, for which air viscosity has an impact in the orbit of the Satellite while still allowing to gather data for a few revolutions.
We will then proceed to compare the position of the Satellite at different times between orbits with and without New Physics ($\alpha\neq 0$ and $\alpha=0$ in eq.~(\ref{eq:BNPotential}), respectively). 

\begin{figure}
    \centering
    \includegraphics[width=0.45\textwidth, trim={2cm 0.5cm 0cm 0cm}]{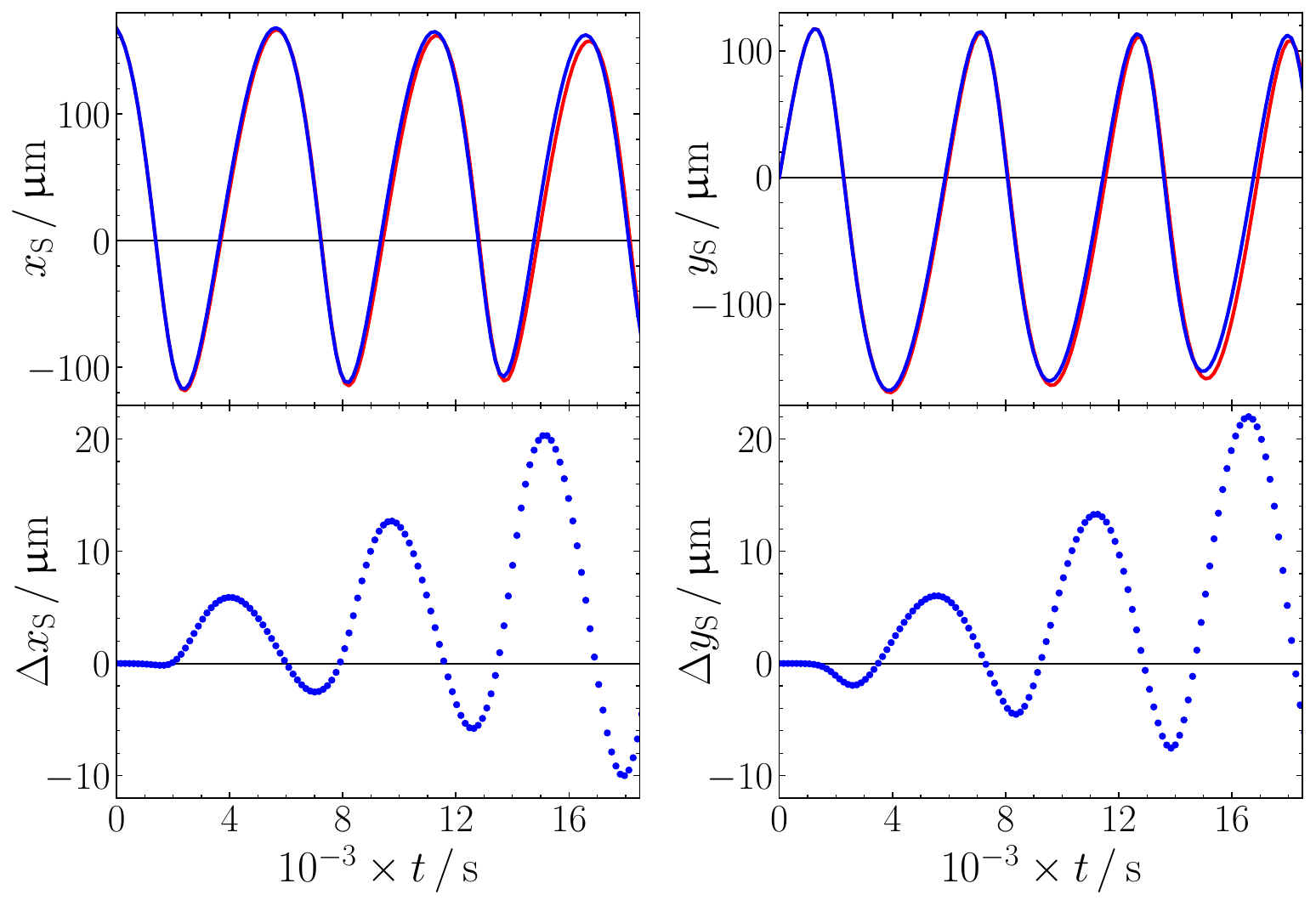}
    \caption{\it Upper panels: the projection of the Satellite position onto the $x$-axis (left) and the $y$-axis (right) as a function of time in seconds. Red lines refer to the Newtonian orbit and blue lines to the New Physics orbit for the particular choice $\lambda = 16$ $\upmu$\textup{m} and $\alpha = 1$ and initial conditions as in {\bf Case 3}. Nuisance parameters have been fixed to $Q_2 = 10^{-3}$, $Q_3 = 0.30$ $\upmu$\textup{m} and $Q_4 = 1$ $\upmu$\textup{m}$^2$, and the air pressure to $p=2\times10^{-5}$ \textup{mbar}. Lower panels:
    the offset between Newtonian and Beyond Newtonian orbits in the $x$-axis (left) and $y$-axis (right) as a function of time. The gray-shaded region represents the data taking window, as explained in the main text.
    }
    \label{fig:datataking}
\end{figure}

In the upper panels of Fig.~\ref{fig:datataking} we compare the $\alpha=0$ trajectory (red, thin lines) with the case with New Physics (blue, thick lines) as a function of time, using initial conditions as in \textbf{Case 3}. The horizontal axis gives the time at which a certain position is reached in seconds, and the vertical axis shows the projection
of the position of the Satellite onto the $x$-axis, $x_{\rm S}(t)$ (left panel) and the $y$-axis, $y_{\rm S}(t)$ (right panel), respectively, with the position given in micrometers. In both cases, the same set of nuisance parameters as in Fig.~\ref{fig:Limits2} (left panel) has been considered: $Q_2 = 10^{-3}$, $Q_3 = 0.3$ $\upmu$\textup{m} and $Q_4 = 1$ $\upmu$\textup{m}$^2$, as well as $p=2\times 10^{-5}$ mbar.
The New Physics orbit has been computed for a reference choice $\alpha = 1$ and $\lambda = 16$ $\upmu$\textup{m}, but similar results can be obtained for different
values in the ($\lambda,\alpha$) plane. We observe that during the first 3000 s, a lapse of time corresponding approximately to the first half revolution, the two trajectories mostly
coincide. In the lower panels we show the offset between the expected position in the $x$- or the $y$-axis in the absence and presence of New Physics. We can see that the offsets between the two orbits are very
small during the first half revolution ($\mathrm{\Delta} x(t) \sim \mathrm{\Delta} y(t) < 1$ $\upmu$\textup{m}), whereas starting from that point they increase significantly, becoming larger than 10 $\upmu$\textup{m} after a few revolutions. It is only after around half a revolution that the Satellite gets close enough to the Planet to effectively feel the effect of the New Physics, leading to higher discrepancies in the orbits from that point on. Notice that the expected error on the position of the center-of-mass of the Satellite is $\sigma \sim 1$ $\upmu$\textup{m}, and therefore as soon as the offset start to grow we will be able 
to detect deviations from the Newtonian orbit with a very good precision. 

\begin{figure}
    \centering
    \includegraphics[width=0.45\textwidth, trim={1.0cm 0.5cm 0cm 0cm}]{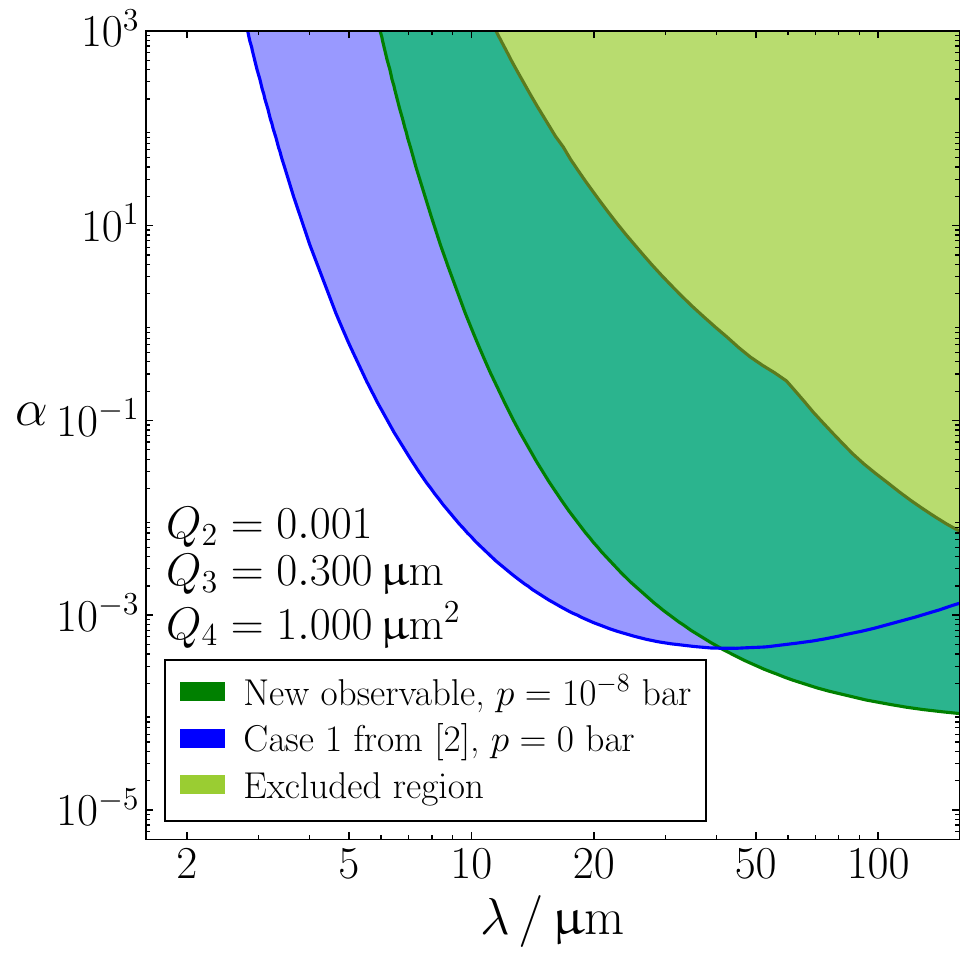}
    \caption{\it 
    Dark-green shaded area: the 95 \% CL sensitivity of the optimized experimental setup in the ($\lambda,\alpha$) plane with air viscosity corresponding to $p = 10^{-5}$ \textup{mbar}, using the $\chi^2$ in eq.~(\ref{eq:finalchi2}), for \textbf{Case 3} intial conditions.  
    Blue-shaded area: the 95 \% CL sensitivity of the setup considered in Ref.~\cite{Baeza-Ballesteros:2021tha} for \textbf{Case 1} initial conditions.
    The light-green shaded region represents the present experimental bounds from Ref.~\cite{Lee:2020zjt}. 
    }
    \label{fig:newsensitivity}
\end{figure}

This is shown in Fig.~\ref{fig:newsensitivity}, where we present our final results for the sensitivity of the newly optimized setup (dark-green region), compared with the present bounds (light-green region) and with the sensitivity of the setup that was optimized in Ref.~\cite{Baeza-Ballesteros:2021tha} (blue region), where the air viscosity was not taken into account. The main difference between the present results
and our previous ones is that we have changed the observable used to detect
New Physics (in addition to a re-optimization of the Planet and Satellite masses, radii and of the initial conditions). In the previous papers, we adopted a very conservative approach, 
in which the data acquisition system only measured the time needed for
the Satellite to perform a full $2\pi$-revolution around the Planet. 
After $N_{\rm rev} = 30$ revolutions, we collected a set of measurements, ${T^1_{\rm BN}(\lambda,\alpha), \dots, T^{N_{\rm rev}}_{\rm BN}(\lambda,\alpha)}$,  and computed the $\chi^2$ between the case with and without New Physics by comparing the maximum variation of the measured periods, $T^\text{max}=\max_{i,j< N}|T^i-T^j|$, in both cases,  with a conservative error $\sigma_{\rm T} = 1$ s. In order to perform the full measurement, 
we therefore needed stable experimental conditions for approximately 
$30 \, T_{\rm N}$, a time span approaching 60 hours for the {\bf Case 1} setup.
Taking advantage of the new data acquisition system (contitued by four Hamamatsu C14440-20UP cameras), in this paper we completely change our
observable. Instead of taking time measurements at a fixed point, we
perform a position measurement at fixed time intervals, with an expected
error on the position of the center-of-mass of the Satellite $\sigma_{x,y} = 1$ $\upmu$\textup{m} both in the $x$- and in the $y$-axis. Considering the position 
along the $x$- and $y$-axis as independent measurements, and taking 
$N_{\rm pos} = 150$ measurements at intervals of 125 s,
 we compare the New Physics to the $\alpha=0$ case,
using the following $\chi^2$:
\begin{eqnarray}
\label{eq:finalchi2}
\chi^2 (\lambda,\alpha) &=& 
\sum_{i=1}^{N_{\rm pos}} \frac{
\left [ x_{\rm BN} (t_i; \lambda, \alpha) - x_{\rm N} (t_i) \right ]^2}{\sigma_x^2} \nonumber \\
&+& 
\sum_{i=1}^{N_{\rm pos}} \frac{
\left [ y_{\rm BN} (t_i; \lambda, \alpha) - y_{\rm N} (t_i) \right ]^2}{\sigma_y^2} \, .
\end{eqnarray}
We note that here we are considering all points independently of the distance between the two bodies, although the two may not be distinguishable for very small separations, as discussed in Sec.~\ref{sec:camera}. Discarding such points from this study has a negligible effect on the presented results.

The present bounds from Ref.~\cite{Lee:2020zjt} are represented by the 
light-green shaded area. On the other hand, the 95 \% CL sensitivity of the setup optimized in vacuum in Ref.~\cite{Baeza-Ballesteros:2021tha} for \textbf{Case 1} initial conditions is represented
by the blue-shaded region. Eventually, the 95~\% CL sensitivity for the setup optimized in presence of air corresponding to $p=10^{-5}$ mbar is shown by the dark-green shaded area, for \textbf{Case 3} initial conditions.
It can be seen that, although we lose some sensitivity compared to the ideal case from Ref.~\cite{Baeza-Ballesteros:2021tha} in the region of large $\alpha$, we still hugely increase the sensitivity with respect to 
present bounds. This is achieved in the presence of air (with the requirement
of a non-extreme air pumping) and with a measuring time of only $\sim 5$ h. In addition, thanks to the newly proposed observable, we are able to explore the interesting region of very low $\alpha$, down to $\alpha \sim 10^{-4}$ for $\lambda \ge 50$ $\upmu$\textup{m}. 

\section{Conclusions}
\label{sec:concl}

Testing Newton's $1/r^2$ law at microscopic distances is an important tool to constrain proposals that have been advanced so far to tame some of the many Standard Model open issues. For example, if new flat spatial dimensions are added to the standard three we are used to (so to solve the hierarchy problem), below some characteristic length scale to be determined, gravity should propagate into a space-time with larger dimensionality with respect to the one usually considered, thus resulting in a dependence of the gravitational force with the distance $r$ between two bodies different from $1/r^2$. Other interesting modification of the gravitational attraction 
at short distances may be motivated by so-called fifth-force models, or quintessence, typically advanced to explain either Dark Matter or Dark Energy observations. With similar motivations, modified gravity models have been
also proposed, in which the Einstein's action is replaced by a more complicated function of the Ricci scalar $R$, or by including terms
proportional to the Ricci tensor $R_{\mu\nu}$, the Riemann tensor 
$R_{\alpha\beta\mu\nu}$ or other rank-2 tensors. In the limit of weak gravitational
field, most of these models can be cast in the form of a Yukawa-like
correction to the Newton's potential, as $\alpha/r\times\exp{(-r/\lambda)}$, with different typical values for the coupling $\alpha$ and for the length scale $\lambda$. 

Present bounds in the $(\lambda,\alpha)$ plane have been obtained measuring the absolute value of the attraction between two bodies (see Refs.~\cite{Adelberger:2009zz,Lee:2020zjt} and references therein). 
At very short distance, however, electrical forces represent an unavoidable background to these measurement. We proposed in Ref.~\cite{Donini:2016kgu} a way to avoid
these backgrounds all-together by measuring the orbit of a microscopic-sized body around a second, larger, one, instead than their mutual absolute attraction. This is because a potential different from $1/r$ will induce precession of the orbit, whereas corrections to the coupling of a $1/r$ potential, such as those coming from electrostatic attraction, may only change the time needed to perform an otherwise closed orbit. A possible experimental setup designed to take advantage of an orbit measurement was sketched. We further studied this possibility in Ref.~\cite{Baeza-Ballesteros:2021tha} by introducing possible nuisance parameters (such as General Relativity or Casimir corrections) that may induce precession of the orbit and must be constrained in order to assess the real sensitivity of the setup. We were able to show that such a setup should be capable
of improving present bounds  by one order of magnitude for some region
of the $(\lambda,\alpha)$ plane. 

In this paper, we go one step forward in designing a realistic experimental setup to measure the orbit of a microscopical-sized planetary system. We take into account the presence of air, that will 
affect the orbit, albeit in a different way from the nuisance parameters included before. We have studied the amount of air dilution required
to preserve the goal sensitivity, finding 
that depleting the air into the Lab volume by a factor $10^{-9}-10^{-8}$
with respect to the standard atmospheric pressure suffices
to preserve the sensitivity of the setup proposed in Refs.~\cite{Donini:2016kgu,Baeza-Ballesteros:2021tha}. We have added
to our setup the corresponding prescriptions for air pumping and started to
characterize the Lab working conditions. Taking advantage of the need for
a more realistic design of the setup, we also introduce a different data taking approach. In Refs.~\cite{Donini:2016kgu,Baeza-Ballesteros:2021tha} we conservatively considered to measure the time
required for the Satellite to perform a full $2 \pi$-revolution around
the Planet, $T^i_{\rm BN}$, comparing with the expected fixed (or almost
fixed, depending on the nuisance parameters $Q_i$) Newtonian period $T_{\rm N}$. Now we propose to use 4 identical 
Hamamatsu C14440-20UP cameras located symmetrically 
with respect to the Lab volume to take repeated measurements of absolute the position of the Satellite
along its orbit. Whilst in the former approach we need a significant number of
revolutions to maximize the sensitivity (we considered typically
30 full revolutions), in the latter case we have found that after three or four
full revolutions the amount of accumulated data is enough to improve the current bounds on the $(\lambda,\alpha)$ plane. This means that, whereas within the former approach
we need stable working conditions for a time span of several dozens of hours,
with the new data gathering approach we need to keep a stable setup only for a few hours, thus simplifying significantly the feasibility of the experiment. 
Our final sensitivity typically range down
to a few microns for large $\alpha$, $\alpha \sim 10^3$, and to $\alpha > 10^{-3}$ for $\lambda \sim 100$ $\upmu$\textup{m}. Notice that present bounds for these parameters are $\lambda \lesssim 10$ $\upmu$\textup{m} for $\alpha \sim 10^3$ and $\alpha \lesssim 5 \times 10^{-2}$ for $\lambda \sim 100$ $\upmu$\textup{m}.

\begin{acknowledgement}
We are grateful with J. J. Gómez-Cadenas for some insightful discussions in the preliminary stages of this work. JBB and AD are partially supported by the EU H2020 research and innovation programme under the MSC grant agreement No 860881-HIDDeN, and the Staff Exchange grant agreement No-101086085-ASYMMETRY, as well as by the Spanish Ministerio de Ciencia e Innovaci\'on project PID2020-113644GB-I00 and by Generalitat Valenciana through the grant PROMETEO/2019/083. JBB is also supported by the Spanish grant FPU19/04326 of MU.   GMT acknowledges financial support from CSIC Research Platform PTI-001 and from the Spanish Ministerio de Ciencia e Innovaci\'on project PID2022-143268NB-I00. FM acknowledges support European Research Council (ERC) under Grant Agreement No. 101039048-GanESS. AS acknowledges support from the European Union's Horizon 2020 research and innovation programme under the MSC grant agreement No 101026628. 
\end{acknowledgement}

\bibliographystyle{h-elsevier}
\bibliography{references}

\end{document}